\documentclass[compsoc,a4paper]{IEEEtran}
\usepackage{changes}
\usepackage{graphicx}
\usepackage{etex}
\usepackage[english]{babel}
\usepackage{amssymb,amsmath,url}
\usepackage{stmaryrd}
\usepackage{ifthen}
\usepackage[normalem]{ulem}
\usepackage{tikz}
\usetikzlibrary{arrows,calc}
\usetikzlibrary{automata,positioning,decorations.pathreplacing}

\usepackage{color}
\usepackage{multirow,multicol}
\usepackage{calc}
\usepackage{array}
\usepackage{msc}
\usepackage{xspace}
\usepackage{centernot}

\usepackage{subfigure}
\usepackage{caption}

\usepackage{enumerate}

\xdefinecolor{DarkGreen}{cmyk}{0.8,0,0.8,0.3}
\definecolor{grayone}{gray}{0.90}
\definecolor{graytwo}{gray}{0.75}
\definecolor{graythree}{gray}{0.75}

\newtheorem{example}{\bf Example}

\begin{document}

\let\v\mathit
\let\f\mathit

\newcommand{\Tamarin}{{\sc Tamarin}\xspace}

\newcommand{\CdsDigest}[1]{\mathsf{digest}(#1)}
\newcommand{\CdsContent}[1]{\mathsf{content}(#1)}
\newcommand{\OdsDigest}[1]{\mathsf{digest}_o(#1)}
\newcommand{\OdsContent}[1]{\mathsf{content}_o(#1)}

\newcommand{\CLog}[1]{\mathsf{Clog}(#1)}
\newcommand{\OLog}[2]{\mathsf{Olog}_{#2}(#1)}
\newcommand{\CDigest}[1]{\mathsf{Cdgt}(#1)}
\newcommand{\ODigest}[2]{\mathsf{Odgt}_{#2}(#1)}

\newcommand{\OLVerifPoAddName}{\mathsf{VerifPoAdd}_O}
\newcommand{\OLVerifPoAbsName}{\mathsf{VerifPoAbs}_O}
\newcommand{\OLVerifPoDName}{\mathsf{VerifPoD}_O}
\newcommand{\OLVerifPoMName}{\mathsf{VerifPoM}_O}
\newcommand{\OLVerifPoPName}{\mathsf{VerifPoP}_O}
\newcommand{\OLVerifPoUAbsName}{\mathsf{VerifPoUAbs}_O}

\newcommand{\OLVerifPoAdd}[4]{\OLVerifPoAddName(#1,#2,#3,#4)}
\newcommand{\OLVerifPoAbs}[3]{\OLVerifPoAbsName(#1,#2,#3)}
\newcommand{\OLVerifPoD}[4]{\OLVerifPoDName(#1,#2,#3,#4)}
\newcommand{\OLVerifPoM}[5]{\OLVerifPoMName(#1,#2,#3,#4,#5)}
\newcommand{\OLVerifPoP}[3]{\OLVerifPoPName(#1,#2,#3)}
\newcommand{\OLVerifPoUAbs}[4]{\OLVerifPoUAbsName(#1,#2,#3,#4)}

\newcommand{\add}{\mathsf{add}}
\newcommand{\del}{\mathsf{del}}
\newcommand{\newlog}{\mathsf{new}}
\newcommand{\blacklist}{\mathsf{bl}}
\newcommand{\beginup}{\mathsf{beg}}
\newcommand{\finup}{\mathsf{end}}
\newcommand{\change}{\mathsf{mod}}

\newcommand{\Rreq}{\leq_r}
\newcommand{\lexA}{\mathop{<_\A}}


\newcommand{\true}{\mathsf{true}}
\newcommand{\false}{\mathsf{false}}


\newcommand{\R}{\mathop{\mathcal{R}}}			
\newcommand{\notR}{\centernot\R}				
\newcommand{\lexico}{\mathop{\mathcal{R}_\ell}}	


\newcommand{\A}{\mathcal{A}}						

\newcommand{\position}[1]{\mathsf{pos}(#1)}			
\newcommand{\nodelbl}[1]{\mathsf{lbl}(#1)}			
\newcommand{\height}[1]{\mathsf{height}(#1)}			
\newcommand{\depth}[2]{\mathsf{depth}_{#2}(#1)}			

\newcommand{\Root}[1]{\mathsf{root}(#1)}			
\newcommand{\Lchild}[2]{\mathsf{Lchild}_{#2}(#1)}		
\newcommand{\Rchild}[2]{\mathsf{Rchild}_{#2}(#1)}		
\newcommand{\Parent}[2]{\mathsf{Parent}_{#2}({#1})}	
\newcommand{\Sib}[2]{\mathsf{Sib}_{#2}({#1})}			

\newcommand{\Order}[1]{\mathop{\mathcal{O}_{#1}}}                       

\newcommand{\dpt}{\mathop{::}}


\newcommand{\Seq}[1]{\mathcal{S}(#1)}                                   
\newcommand{\CT}[1]{\mathcal{CT}(#1)}					

\newcommand{\CTPosVerify}[2]{\mathsf{VerifPos}_C(#1,#2)}
\newcommand{\CTPosOfIndex}[2]{\mathsf{Pos\_of\_Ind}(#1,#2)}
\newcommand{\CTIndexOfPos}[2]{\mathsf{Ind\_of\_Pos}(#1,#2)}

\newcommand{\CTPoP}[2]{(#1,#2)}
\newcommand{\CTComputePoP}[3]{\mathsf{CompPoP_C}(#1,#2,#3)}
\newcommand{\CTVerifPoPName}{\mathsf{VerifPoP}_c}
\newcommand{\CTVerifPoP}[3]{\CTVerifPoPName(#1,#2,#3)}

\newcommand{\CTVerifPoEName}{\mathsf{VerifPoE}_c}
\newcommand{\CTVerifPoE}[3]{\CTVerifPoEName(#1,#2,#3)}

\newcommand{\RandomVerifCT}[4]{\mathsf{Rand\exists}_C(#1,#2,#3,#4)}
\newcommand{\RandomVerifCTName}{\mathsf{Rand\exists}_C}
\newcommand{\RandomVerifAVL}[4]{\mathsf{Rand\exists}_{AVL}(#1,#2,#3,#4)}
\newcommand{\RandomVerifAVLName}{\mathsf{Rand\exists}_{AVL}}

\newcommand{\OLRandomVerifMName}{\mathsf{RandM}_O}
\newcommand{\OLRandomVerifM}[7]{\OLRandomVerifMName(#1,#2,#3,#4,#5,#6,#7)}


\newcommand{\Data}[1]{\mathsf{Data}(#1)}					

\newcommand{\LTPoP}[4]{(#1,#2,#3,#4)}
\newcommand{\LTComputePoP}[4]{\mathsf{CompPoP}_L(#1,#2,#3,#4)}
\newcommand{\LTVerifPoP}[3]{\mathsf{VerifPoP}_L(#1,#2,#3)}

\newcommand{\LTPoAbs}[4]{(#1,#2,#3,#4)}
\newcommand{\LTVerifPoAbs}[3]{\mathsf{VerifPoAbs}_L(#1,#2,#3)}

\newcommand{\lextreeofdata}{\mathsf{lex}}


\newcommand{\factorheight}[2]{\mathsf{fctH}_{#2}(#1)}
\newcommand{\additionAVL}[2]{\mathsf{add}(#1,#2)}
\newcommand{\additionLexAVL}[2]{\mathsf{add}_{Lex}(#1,#2)}
\newcommand{\modifyLexAVL}[2]{\mathsf{mod}_{Lex}(#1,#2)}

\newcommand{\toLexAVL}{\mathsf{to}_{Lex}}

\newcommand{\AVLPoP}[6]{(#1,#2,#3,#4,#5,#6)}
\newcommand{\AVLVerifData}[1]{\mathsf{VerifData}_{AVL}(#1)}
\newcommand{\AVLVerifPoP}[3]{\mathsf{VerifPoP}_{AVL}(#1,#2,#3)}
\newcommand{\AVLVerifPoM}[5]{\mathsf{VerifPoM}_{AVL}((#1,#2),(#3,#4),#5)}

\newcommand{\AVLVerifPoAdd}[4]{\mathsf{VerifPoAdd}_{AVL}(#1,#2,#3,#4)}
\newcommand{\AVLVerifPoAbs}[3]{\mathsf{VerifPoAbs}_{AVL}(#1,#2,#3)}

\newcommand{\CompH}{\mathsf{CompH}}
\newcommand{\Comph}{\mathsf{Comph}}
\newcommand{\CompfH}{\mathsf{Compf}}


\newcommand{\regex}[1]{Sreg_{#1}}
\newcommand{\Rreg}{\mathop{\mathcal{R}_{reg}}}
\newcommand{\regexVerifname}{\mathsf{VerifReg}}
\newcommand{\regexVerif}[2]{\regexVerifname(#1,#2)}


\newcommand{\revlist}{{rev_\ell}}
\newcommand{\addlist}{{add_\ell}}
\newcommand{\sizelist}{{s_\ell}}


\newcommand{\certset}{\mathcal{C}}
\newcommand{\Rcert}{\mathop{\mathcal{R}_{\certset}}}
\newcommand{\key}{\mathsf{key}}
\newcommand{\id}{\mathsf{id}}
\newcommand{\hashset}{\mathcal{H}}

\newcommand{\addition}{\mathsf{add}}
\newcommand{\revoke}{\mathsf{rev}}
\newcommand{\requestset}{\mathsf{Req}}

\newcommand{\ProofCons}[4]{(#1,#2,#3,#4)}

\newcommand{\domname}{\mathit{domain}}
\newcommand{\logname}{\mathit{nlog}}
\newcommand{\cert}{\mathit{cert}}
\newcommand{\hlog}{h}
\newcommand{\reqconsistency}{\mathsf{req}_{\mathsf{ext}}}
\newcommand{\reqextension}{\mathsf{req}_{\mathsf{cons}}}
\newcommand{\reqabsence}{\mathsf{req}_{\mathsf{abs}}}
\newcommand{\registration}{\mathsf{reg}}
\newcommand{\revocation}{\mathsf{revoc}}
\newcommand{\validation}{\mathsf{valid}}

\newcommand{\act}{\mathsf{act}}
\newcommand{\Blist}{\mathsf{bl}}


\newcommand{\size}{\mathsf{size}}
\newcommand{\sizesign}{\mathsf{size}_{\mathsf{sign}}}
\newcommand{\sizehash}{\mathsf{size}_{\mathsf{hash}}}
\newcommand{\sizecert}{\mathsf{size}_{\mathsf{cert}}}
\newcommand{\sizedata}{\mathsf{size}_{\mathsf{data}}}

\newcommand{\hfun}{\mathsf{h}}					
\newcommand{\Null}{\mathsf{null}}

\newcommand{\head}{\mathsf{head}}
\newcommand{\tail}{\mathsf{tail}}
\newcommand{\cons}{\mathsf{::}}
\newcommand{\pair}[2]{\langle#1,#2\rangle}		
\newcommand{\proj}{\mathsf{proj}}				
\newcommand{\senc}{\mathsf{senc}}			
\newcommand{\sdec}{\mathsf{sdec}}			
\newcommand{\aenc}{\mathsf{aenc}}			
\newcommand{\adec}{\mathsf{adec}}			
\newcommand{\arenc}{\mathsf{aenc}_r}			
\newcommand{\ardec}{\mathsf{adec}_r}			
\newcommand{\pk}{\mathsf{pk}}				
\newcommand{\sign}{\mathsf{sign}}				
\newcommand{\checksign}{\mathsf{check}}		
\newcommand{\get}{\mathsf{getmsg}}
\newcommand{\vk}{\mathsf{vk}}				
\newcommand{\h}{\mathsf{h}}					
\newcommand{\mac}{\mathsf{mac}}				

\newcommand{\checkinlog}{\mathsf{checklog}}
\newcommand{\request}{\mathsf{req}}
\newcommand{\register}{\mathsf{reg}}
\newcommand{\Register}{\mathsf{Register}}
\newcommand{\Request}{\mathsf{Request}}
\newcommand{\CA}{\mathsf{CA}}
\newcommand{\logo}{\mathsf{log}}
\newcommand{\inter}{\mathsf{inter}}
\newcommand{\cache}{\mathsf{cache}}


\newcommand{\master}{\mathsf{master}}
\newcommand{\tls}{\mathsf{tls}}
\newcommand{\reg}{\mathsf{reg}}
\newcommand{\rev}{\mathsf{rev}}
\newcommand{\upadd}{\mathsf{upadd}}
\newcommand{\updel}{\mathsf{updel}}

\newcommand{\pkset}{\mathcal{PK}}
\newcommand{\digestset}{\mathcal{D}gt}
\newcommand{\chfune}[2]{\mathsf{dg_{#1,#2}}}
\newcommand{\ckfune}[1]{\mathsf{keys}_{#1}}

\definecolor{light-gray}{gray}{0.95}
\definecolor{medium-gray}{gray}{0.75}

\title{DTKI: a new formalized PKI with verifiable trusted parties}

\author{ \IEEEauthorblockN{Jiangshan Yu\thanks{Corresponding Author:
      Jiangshan Yu. Email: j.yu.research@gmail.com.\newline This paper is
      published at {\it The Computer Journal}, Vol.59 No.11,
      pp. 1695-1713, 2016}\IEEEauthorrefmark{1}, Vincent
    Cheval\IEEEauthorrefmark{2}, %
    and Mark Ryan\IEEEauthorrefmark{1}\\[3ex]}
  \IEEEauthorblockA{\IEEEauthorrefmark{1}University of Birmingham,
    United Kingdom}\\ \IEEEauthorblockA{\IEEEauthorrefmark{2}LORIA,
    CNRS, France}}



\maketitle
\begin{abstract}
  The security of public key validation protocols for web-based
  applications has recently attracted attention because of weaknesses
  in the certificate authority model, and consequent attacks.

  Recent proposals using public logs have succeeded in making
  certificate management more transparent and verifiable. However,
  those proposals involve a fixed set of authorities. This means an
  oligopoly is created. Another problem with current log-based system
  is their heavy reliance on trusted parties that monitor the logs.

  We propose a distributed transparent key infrastructure (DTKI),
  which greatly reduces the oligopoly of service providers and allows
  verification of the behaviour of trusted parties. In addition, this
  paper formalises the public log data structure and provides a formal
  analysis of the security that DTKI guarantees.
\end{abstract}
\begin{IEEEkeywords}
  PKI, SSL, TLS, key distribution, certificate, transparency, trust,
  formal verification.
\end{IEEEkeywords}

\section{Introduction}
The security of web-based applications such as e-commerce and web-mail
depends on the ability of a user's browser to obtain authentic copies
of the public keys for the application website. For example, suppose a
user wishes to log in to her bank account through her web browser. The
web session will be secured by the public key of the bank. If the
user's web browser accepts an inauthentic public key for the bank,
then the traffic (including log-in credentials) can be intercepted and
manipulated by an attacker.

The authenticity of keys is assured at present by \emph{certificate
  authorities} (CAs). In the given example, the browser is presented
with a public key certificate for the bank, which is intended to be
unforgeable evidence that the given public key is the correct one for
the bank. The certificate is digitally signed by a CA. The user's
browser is pre-configured to accept certificates from certain known
CAs. A typical installation of Firefox has about 100 root certificates
in its database.

Unfortunately, numerous problems with the current CA model have been
identified. Firstly, CAs must be assumed to be trustworthy. If a CA is
dishonest or compromised, it may issue certificates asserting the
authenticity of fake keys; those keys could be created by an attacker
or by the CA itself. Secondly, the assumption of honesty does not
scale up very well. As already mentioned, a browser typically has
hundreds of CAs registered in it, and the user cannot be expected to
have evaluated the trustworthiness and security of all of them. This
fact has been exploited by attackers
\cite{cert-iranians,cert-trustwave,cert-ms-verisign,news4,news5,news6}.
In 2011, two CAs were compromised: Comodo \cite{cert-comodo} and
DigiNotar \cite{cert-diginotar}. In both cases, certificates for
high-profile sites were illegitimately obtained, and in the second
case, reportedly used in a \emph{man in the middle} (MITM) attack
\cite{cert-iranian-guardian}.

\subsection*{Proposed solutions}

Several interesting solutions have been proposed to address these
problems. For a comprehensive survey, see \cite{jeremyclark:s+p2013}.

\smallskip

Key pinning mitigates the problem of untrustworthy CAs, by defining in
the client browser the parameters concerning the set of CAs that are
considered entitled to certify the key for a given domain
\cite{Langley:pinning,cert-TACK}. However, scalability is a challenge
for key pinning.

Crowd-sourcing techniques have been proposed in order
to detect untrustworthy CAs, by enabling a browser to obtain warnings
if the received certificates are different from those that other
people are being offered \cite{P08, cert-ssl-obs, cert-doublecheck,
  cert-icsi-notary, observatory, CP, Certlock, Con11}. Crowd-sourcing
techniques have solved some CA-based problems. However, the technique
cannot distinguish between attacks and authentic certificate updates,
and may also suffer from an initial unavailability period.

Solutions for revocation management of certificates have also been
proposed; they mostly involve periodically pushing revocation lists to
browsers, in order to remove the need for on-the-fly revocation
checking \cite{cert-rivest-crl, Langley:revocation}. However, these
solutions create a window during which the browser's revocation lists
are out of date until the next push.

\smallskip

More recently, solutions involving public append-only logs have been
proposed. We consider the leading proposals here.

\bigskip

\paragraph*{\textbf{Public log-based systems}}

\emph{Sovereign Keys} (SK) \cite{SovereignKeys} aims to get rid of
browser certificate warnings, by allowing domain owners to establish a
long term (``sovereign'') key and by providing a mechanism by which a
browser can hard-fail if it doesn't succeed in establishing security
via that key. The sovereign key is used to cross-sign operational TLS
\cite{rfc5246,rfc6176} keys, and it is stored in an append-only log on
a ``time-line server'', which is abundantly mirrored. However, in SK,
internet users and domain owners have to trust mirrors of time-line
servers, as SK does not enable mirrors to provide efficient verifiable
proofs that the received certificate is indeed included in the
append-only log.

\medskip

\emph{Certificate transparency} (CT) \cite{rfc6962} is a technique
proposed by Google that aims to efficiently detect fake public key
certificates issued by corrupted certificate authorities, by making
certificate issuance transparent. They improved the idea of SK by
using append-only Merkle tree to organise the append-only log. This
enables the log maintainer to provide two types of verifiable
cryptographic proofs: (a) a proof that the log contains a given
certificate, and (b) a proof that a snapshot of the log is an
extension of another snapshot (\emph{i.e.}, only appends have taken
place between the two snapshots). The time and size for proof
generation and verification are logarithmic in the number of
certificates recorded in the log. Domain owners can obtain the proof
that their certificates are recorded in the log, and provide the proof
together with the certificate to their clients, so the clients can get
a guarantee that the received certificate is recorded in the log.

\medskip

\emph{Accountable key infrastructure} (AKI) \cite{AKI} also uses
public logs to make certificate management more transparent. By using
a data structure that is based on lexicographic ordering rather than
chronological ordering, they solve the problem of key revocations in
the log. In addition, AKI uses the ``checks-and-balances'' idea that
allows parties to monitor each other's misbehaviour. So AKI limits the
requirement to trust any party. Moreover, AKI prevents attacks that
use fake certificates rather than merely detecting such attacks (as in
CT). However, as a result, AKI needs a strong assumption --- namely,
CAs, public log maintainers, and validators do not collude together
--- and heavily relies on third parties called validators to ensure
that the log is maintained without improper modifications.


\medskip

\emph{Certificate issuance and revocation transparency} (CIRT)
\cite{RyanNDSS14} is a proposal for managing certificates for
end-to-end encrypted email. It proposes an idea to address the
revocation problem left open by CT, and the trusted party problem of
AKI. It collects ideas from both CT and AKI to provide transparent key
revocation, and reduces reliance on trusted parties by designing the
monitoring role so that it can be distributed among user
browsers. However, CIRT can only detect attacks that use fake
certificates; it cannot prevent them. In addition, since CIRT was
proposed for email applications, it does not support the multiplicity
of log maintainers that would be required for web
certificates.

\medskip

\emph{Attack Resilient Public-Key Infrastructure} (ARPKI) \cite{ARPKI}
is an improvement on AKI. In ARPKI, a client can designate $n$ service
providers (e.g. CAs and log maintainers), and only needs to contact
one CA to register her certificate. Each of the designated service
providers will monitor the behaviour of other designated service
providers. As a result, ARPKI prevents attacks even when $n-1$ service
providers are colluding together, whereas in AKI, an adversary who
successfully compromises two out of three designated service providers
can successfully launch attacks \cite{ARPKI}. In addition, the
security property of ARPKI is proved by using a protocol verification
tool called Tamarin prover \cite{Tamarin}. The weakness of ARPKI is
that all $n$ designated service providers have to be involved in all
the processes (i.e. certificate registration, confirmation, and
update), which would cause considerable extra latencies and the delay
of client connections.

\bigskip

In public log-based systems, efforts have been made to integrate
\emph{revocation management} with the certificate auditing. CT
introduced revocation transparency (RT) \cite{RT12} to deal with
certificate revocation management; and in AKI and ARPKI, the public
log only stores currently valid certificates (revoked certificates are
purged from the log). However, the revocation checking processes in
both RT and A(RP)KI are linear in the number of issued certificates
making it inefficient. CIRT allows efficient proofs of non-revocation,
but it does not scale to multiple logs which are required for web
certificates.

\subsection*{Remaining problems}
A foundational issue is the problem of \emph{oligopoly}. The
present-day certificate authority model requires that the set of
global certificate authorities is fixed and known to every browser,
which implies an oligopoly. Currently, the majority of CAs in browsers
are organisations based in the USA, and it is hard to become a
browser-accepted CA because of the strong trust assumption that it
implies. This means that a Russian bank operating in Russia and
serving Russian citizens living in Russia has to use an American CA
for their public key. This cannot be considered satisfactory in the
presence of mutual distrust between nations regarding cybersecurity
and citizen surveillance, and also trade sanctions which may prevent
the USA offering services (such as CA services) to certain other
countries.

None of the previously discussed public log-based systems address this
issue. In each of those solutions, the set of log maintainers (and
where applicable, time-line servers, validators, etc.)  is assumed to
be known by the browsers, and this puts a high threshold on the
requirements to become a log maintainer (or validator,
etc.). Moreover, none of them solve the problem that a multiplicity of
log maintainers reduces the usefulness of transparency, since a domain
owner has to check each log maintainer to see if it has mis-issued
certificates. This can't work if there is a large number of log
maintainers operating in different geographical regions, each one of
which has to be checked by every domain owner.

\medskip

A second issue is the requirement of trusted parties. Currently, all
existing proposals have to rely on some sort of trusted parties or at
least assume that not all parties are colluding together. However, a
strong adversary (e.g. a government agency) might be able to control
all service providers (used by a given client) in a system.

\medskip

A third foundational issue of a different nature is that of analysis
and correctness. SK, CT, AKI and CIRT are large and complex protocols
involving sophisticated data structures, but none of them have been
subjected to rigorous analysis. It is well-known that security
protocols are notoriously difficult to get right, and the only way to
avoid this is with systematic verification. For example, attacks on
AKI and CIRT have been identified in \cite{ARPKI} and in the appendix
of our technical report \cite{Report}, respectively. The flaws may be
easily fixed, but only once they have been identified. It is therefore
imperative to verify this kind of complex protocol. ARPKI is the first
formally verified log-based PKI system. However, they used several
abstractions during modelling in the Tamarin prover. For example, they
represent the underlying log structure (a Merkle tree) as a
list. However, in systems like CIRT and this paper with more complex
data structures, it is important to have a formalised data structure
and its properties to prove the security claim. The formalisation of
complex data structures and their properties in the log-based PKI
systems is a remaining problem.
\medskip

The last problem is the management of certificate revocation. As
explained previously, existing solutions for managing certificate
revocation (e.g. CRL, OCSP, RT) are still unsatisfactory.

\subsection*{This paper}

We propose a new public log-based architecture for managing web
certificates, called \emph{Distributed Transparent Key Infrastructure}
(DTKI), with the following contributions.
\begin{itemize}
\item We identify \emph{anti-oligopoly} as an important property for
  web certificate management which has hitherto not received
  attention.

\item Compared to its predecessors, DTKI is the first system to have
  all desired features --- it minimises the presence of oligopoly,
  prevents attacks that use fake certificates, provides a way to
  manage certificate revocation, verifies output from trusted parties,
  and is secure even if all service providers (e.g. CAs and log
  maintainers) collude together (see Section \ref{sec:security} for
  our security statement). A comparison of the properties of different
  log-based systems can be found in Section \ref{sec:comparison}.

\item We provide formal machine-checked verification of its core
  security property using the Tamarin prover. In addition, we
  formalise the data structures needed for transparent public logs,
  and provide rigorous proofs of their properties.
\end{itemize}

\section{Overview of DTKI}
\label{sec:overview}

Distributed Transparent Key Infrastructure (DTKI) is an infrastructure
for managing keys and certificates on the web in a way which is
\emph{transparent}, minimises \emph{oligopoly}, and allows
verification of the behaviour of trusted parties. In DTKI, we mainly
have the following agents:

\emph{Certificate log maintainers (CLM):} A CLM maintains a database
of all valid and invalid (e.g. expired or revoked) certificates for a
particular set of domains for which it is responsible. It commits to
digests of its log, and provides efficient proofs of presence and
absence of certificates in the log. 
CLMs behave transparently and their actions can be verified.

\emph{A mapping log maintainer (MLM):} To minimise oligopoly, DTKI
does not fix the set of certificate logs. The MLM maintains
association between certificate logs and the domains they are
responsible for. It also commits to digests of the log, and provides
efficient proof of current association, and behaves
transparently. Clients of the MLM are not required to blindly trust the
MLM, because they can efficiently verify the obtained associations.

The MLM has a strategic role of determining the authorised CLMs, and the
mapping log to be maintained rarely changes; therefore it can be
easily governed by an international panel. In practice, ICANN is a
possible party to be given the responsibility to run the MLM.

\emph{Users and their browsers:} They query the MLM, and obtain and
verify the proofs about the mapping of top-level domains (TLDs) to
CLMs. They query CLMs and obtain and verify proofs about certificates.

\emph{Mirrors:} Mirrors are servers that maintain a full copy of the
mapping log and certificate logs respectively downloaded from the MLM
and corresponding CLMs, and the corresponding digest of the log signed
by the log maintainer. In other words, mirrors are distributed copies
of logs. Anyone (e.g. ISPs, CLMs, CAs, domain owners) can be a
mirror. Unlike in SK, mirrors are not required to be trusted in DTKI,
because they give a proof for every association that they send to
their clients. The proof is associated to the digest of the MLM.

\emph{Certificate authorities (CA):} They check the identity of domain
owners, and create certificates for the domain owners' keys. However,
in contrast with today's CAs, the ability of CAs in DTKI is limited
since the issuance of a certificate from a CA is not enough to
convince web browsers to accept the certificate (proof of presence in
the relevant CLM is also needed).

\medskip

In DTKI, each domain owner has two types of certificate, namely TLS
certificate and master certificate. Domain owners can have different
TLS certificates but can only have one master certificate. A TLS
certificate contains the public key of a domain server for a TLS
connection, whereas the master certificate contains a public key,
called ``master verification key''. The corresponding secret key of
the master certificate is called ``master signing key''. Similar to
the ``sovereign key'' in SK \cite{SovereignKeys}, the master signing
key is only used to validate a TLS certificate (of the same subject)
by issuing a signature on it. This limits the ability of certificate
authorities since without having a valid signature (issued by using
the master signing key), the TLS certificate will not be
accepted. Hence, the TLS secret key is the one for daily use; and the
master signing key is rarely used. It will only be used for validating
a new certificate, or revoking an existing certificate. We assume that
domain owners can take care of their master signing key, as a master
signing key can be kept offline, and is rarely used.

After a domain owner obtains a master certificate or a TLS certificate from a
CA, he needs to make a registration request to the corresponding CLM to publish
the certificate into the log. To do so, the domain owner signs the certificate
using the master signing key, and submits the signed certificate to a CLM
determined (typically based on the top-level domain) by the MLM. The CLM checks
the signature, and accepts the certificate by adding it to the certificate log
if the signature is valid. The process of revoking a certificate is handled
similarly to the process of registering a certificate in the log.

When establishing a secure connection with a domain server, the
browser receives a corresponding certificate and proofs from a mirror
of the MLM and a CLM, and verifies the certificate, the proof that the
certificate is valid and recorded in the certificate log, and proof
that this certificate log is authorised to manage certificates for the
domain. Users and their browsers should only accept a certificate if
the certificate is issued by a CA, and validated by the domain owner,
and current in the certificate log.

Fake master certificates or TLS certificates can be easily detected by
the domain owner, because the attacker will have had to insert such
fake certificates into the log (in order to be accepted by browsers),
and is thus visible to the domain owner.

\medskip

Rather than relying solely on trusted monitors to verify the
healthiness of logs and the relations between logs, DTKI uses a
crowdsourcing-like way to ensure the integrity of the log and the
relations between mapping log and a certificate log, and between
certificate logs. In particular, the monitoring work in DTKI can be
broken into independent little pieces, and thus can be done by
distributing the pieces to users' browsers. In this way, users'
browsers can perform randomly-chosen pieces of the monitoring role in
the background (e.g. once a day).  Thus, web users can collectively
monitor the integrity of the logs. We envisage parameters in
browsers allowing users to control how that works.

To avoid the case that attackers create a ``bubble'' (i.e. an isolated
environment) around a victim, we share the same assumption as other
existing protocols (e.g. CT and CIRT) -- we assume that gossip
protocols \cite{gossip} are used to disseminate digests of the
log. So, users of logs can detect if a log maintainer shows different
versions of the log to different sets of users. Since log maintainers
sign and time-stamp their digests, a log maintainer that issues
inconsistent digests can be held accountable.
\section{The public log}
\label{sec: public log}

DTKI uses append-only logs to record all requests processed by the log
maintainer, and allows log maintainers to efficiently generate some
proofs that can be efficiently verified. These proofs mainly include
that some data (e.g. a certificate or a revocation request) has or has
not been added to the log; and that a log is extended from a previous
version.

So, the log maintainer's behaviour is transparent to the public, and
the public is not required to blindly trust log maintainers. Public
log data structures have been widely studied \cite{Merkle87, PAD01,
  Maniatis03, Aydan07, SovereignKeys,rfc6962,RyanNDSS14}. To the best
of our knowledge, no single data structure can provide all proofs
required by DTKI. We adopt and extend the idea of CIRT log structure
\cite{RyanNDSS14} which makes use of two data structures to provide
all the kinds of proofs needed for DTKI.

This section presents the intuition of two abstract data structures
encapsulating the desired properties, then introduces how to use the
data structures to construct our public logs in a concrete manner by
extending the CIRT data structure. The formalisation of our abstract
data structures, log structures, and their properties, and our
detailed implementation, are presented in our technical report
\cite{Report}. We also present some examples of the data structures
there.

\begin{table}
\setlength\tabcolsep{12pt}
  \centering
 \small
 \begin{tabular}{>{\raggedright}p{.075\textwidth}
     >{\raggedright\arraybackslash}p{.35\textwidth}
     @{}}
   Function & Output\\
   \hline\\[-1ex]
   \multicolumn{2}{l}{\textbf{Chronological Data Structure}}\\[1ex]
   $\mathsf{digest}$ & given input a sequence $S$ of data, it outputs the digest of
   sequence $S$ of data organised by using chronological data
   structure\\
   $\CTVerifPoPName$ & given input $(\CdsDigest{S},d,p)$, it outputs
   a boolean value indicating the verification result of the
   proof $p$ that some data $d$ is included in a set $S$\\
   $\CTVerifPoEName$ & given input $((dg',N'),(dg,N),p)$, it
   outputs a boolean value indicating the verification result of
   the proof $p$ that a sequence of data represented by its
   digest $dg$ and size $N$ is extended from another sequence of
   data represented by digest $dg'$ and size $N'$\\
   &\\[-1ex]
   \multicolumn{2}{l}{\textbf{Ordered Data Structure}}\\[1ex]
   $\mathsf{digest_O}$ & given input a sequence $S$ of data, it outputs the digest of
   sequence $S$ of data organised by using ordered data
   structure\\
   $\OLVerifPoPName$ (resp. $\OLVerifPoAbsName$) & given input $(\OdsDigest{S},d,p)$, it outputs
   a boolean value indicating the verification result of the
   proof $p$ that some data $d$ is (resp. is not) included in a set $S$\\[1ex]
   $\OLVerifPoAddName$ (resp. $\OLVerifPoDName$) & given input
   $(d,dg,dg',p)$, it outputs a boolean value indicating the
   verification result of the proof $p$ that $dg'$ is the digest
   obtained after adding data $d$ into (resp. deleting data $d$ from) the sequence of data
   represented by digest $dg$\\
   $\OLVerifPoMName$ & given input $(d,d',dg,dg',p)$, it outputs
   a boolean value indicating the verification result of the
   proof $p$ that $dg'$ is the digest obtained after replacing
   $d$ with $d'$ in the sequence of data represented by $dg$\\[1ex]
   \hline
  \end{tabular}
  \caption{Some functions supported by the data structures, of size $N$. The full list of operations and functions supported by the data structures, and the detailed properties of the data structures, are formalised in our technical report.}
  \label{tab:operation}
\end{table}

\subsection{Data structures}
Our log makes use of two data structures, namely chronological data
structure and ordered data structure, to provide all the proofs
required by DTKI. We use the notion of \emph{digest} to represent a
unique set of data, such that the size of a digest is a constant. For
example, a digest could be the hash value of a set of data.

A chronological data structure is an append-only data structure,
i.e. only the operation of adding some data is allowed. With a
chronological data structure, for a given sequence $S$ of data of size
$N$ and with digest $dg$, we have $d \in S$ for some data $d$, if and
only if there exists a proof $p$ of size $O(\log(N))$, called the
proof of presence of $d$ in $S$, such that $p$ can be efficiently
verified by using $\CTVerifPoPName$ (see Table \ref{tab:operation});
and for all sequence $S'$ with digest $dg'$ and size $N'<N$, we have
that $S'$ is a prefix of $S$, if and only if there exists a proof $p'$
of size $O(\log(N))$, called the proof of extension of $S$ from $S'$,
such that $p'$ can be efficiently verified by using $\CTVerifPoEName$
(see Table \ref{tab:operation}).


In this way, to verify that some data is included in a sequence of
data stored in a chronological data structure (of size $N$), the
verifier only needs to download the corresponding digest, and the
corresponding proof of presence (with size $O(log(N))$). The
verification of proof of extension is similarly efficient. Possible
implementations are append-only Merkle tree \cite{Merkle87} and
append-only skip list, as proposed in \cite{rfc6962} and
\cite{Maniatis03}, respectively.

With the append-only property, the chronological data structure
enables one to prove that a version of the data structure is an
extension of a previous version. This is useful for our public log
since it enables users to verify the history of a log maintainer's
behaviours.

\medskip

Unfortunately, the chronological data structure does not provide all
desired features. For example, it is very inefficient to verify that
some data (e.g. a revocation request) is not in the chronological data
structure (the cost is $O(N)$, where $N$ is the size of the data
structure). To provide missing features, we need to use the
\emph{ordered data structure}.

An ordered data structure is a data structure allowing one to insert,
delete, and modify stored data. In addition, with an ordered data
structure, for a given sequence $S$ of data of size $N$ and with
digest $dg$, we have $d\in S$ (resp. $d\notin S$) for some data $d$,
if and only if there exists a proof $p$ of size $O(\log(N))$, called
the proof of presence (resp. absence) of $d$ in (resp. not in) $S$,
such that $p$ can be efficiently verified by using $\OLVerifPoPName$
(resp. $\OLVerifPoAbsName$) (see Table \ref{tab:operation}).

Possible implementations of ordered data structure are Merkle tree
which is organised as a binary search tree (as proposed in
\cite{RyanNDSS14}), and authenticated dictionaries \cite{PAD01}.

With an ordered data structure, however, the size of proof that the
current version of the data is extended from a previous version is
$O(N)$. As the chronological data structure and the ordered data
structure have complementary properties, we use the combination of
them to organise our log.


\subsection{Mapping log}
\label{sec:mlog}
To minimise oligopoly, DTKI uses multiple certificate logs, and does
not fix the set of certificate logs and the mapping between domains
and certificate logs. A mapping log is used to record associations
between domain names and certificate log maintainers, and can provide
efficient proofs regarding the current association. It would be rather
inefficient to explicitly associate each domain name to a certificate
log, due to the large number of domains. To efficiently manage the
association, we use a class of simple regular expressions to present a
group of domain names, and record the associations between regular
expressions and certificate logs in the mapping log. For example, the
mapping might include (*$\backslash$.org, Clog$_1$) and
([a-h].*$\backslash$.com, Clog$_1$) to mean that certificate log
maintainer Clog$_1$ deals with domains ending $.org$ and domains
starting with letters from $a$ to $h$ ending $.com$. In our
technical report \cite{Report}, we have formally defined some
constraints on the regular expressions we use, the relations between
them, and how to use random verification to verify that no overlap
between regular expressions exists.

\medskip

\begin{figure}[!htbp]
\includegraphics[width=.5\textwidth]{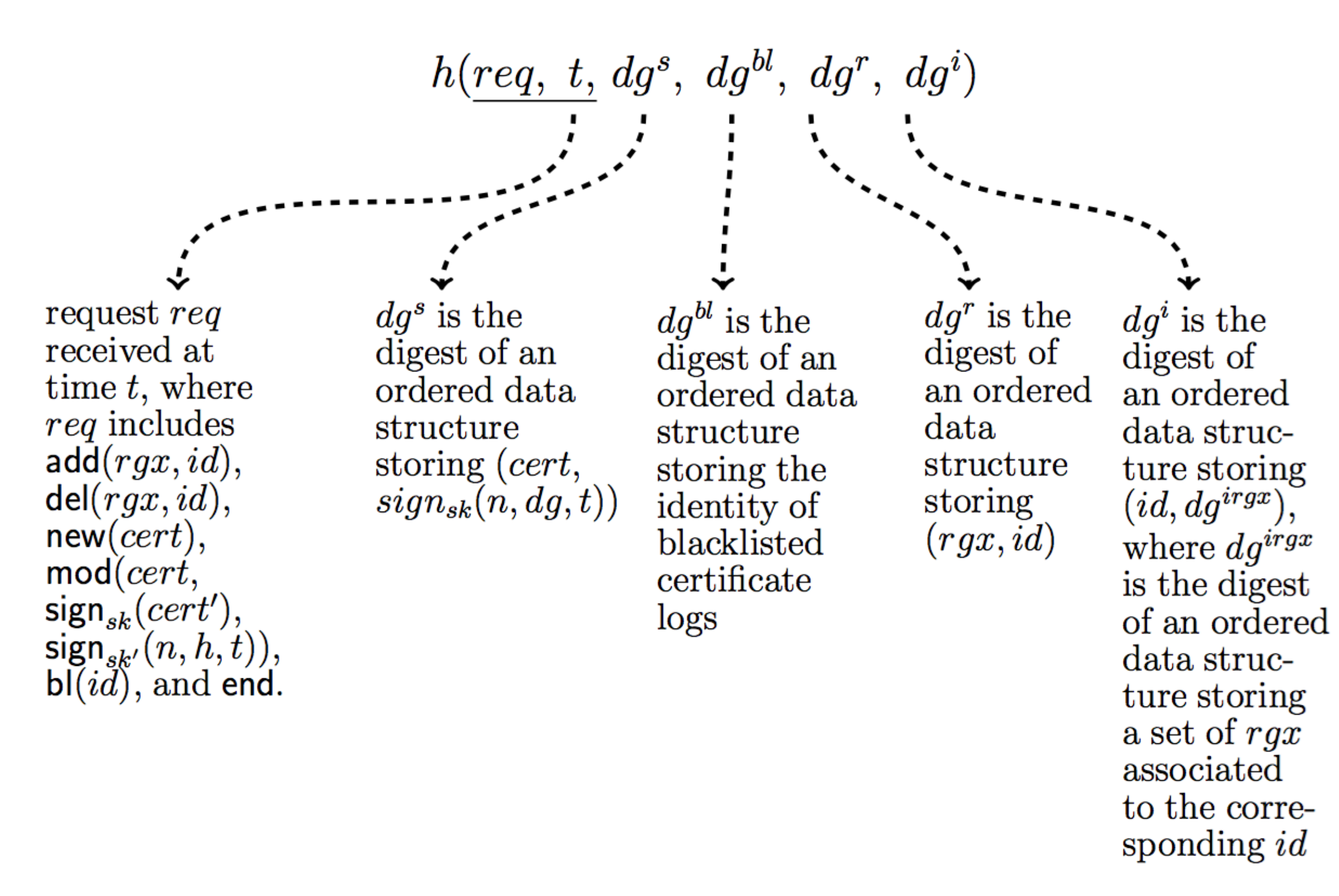}
\caption{A figure representation of the format of each record in the mapping log.
}
  \label{eg:mlog}
\end{figure}

Intuitively, as presented in Figure \ref{eg:mlog}, the mapping log is
organised by using a chronological data structure, and stores received
requests\footnote{ The request includes adding, removing, and
  modifying a certificate log and/or a mapping.}  together with the
request time, and four digests of different ordered data structures
representing the status of the log. Each entry is of the form
\[ \hfun(req,t, dg^s, dg^{bl}, dg^r, dg^i) \]
In the formula, $req$ is the request received by the mapping log at
time $t$; $dg^s$\footnote{We simplified the description here: we
  should say the ordered data structure represented by $dg^s$ stores
  the information, rather than the digest $dg^s$ stores it. We will
  use this simplification through the paper.}  stores information
about CLMs (e.g. the certificate of the CLM, and the current digest of
the certificate log $clog$); $dg^{bl}$ stores the identity of
blacklisted certificate log maintainers; $dg^{r}$ stores the mapping
from a regular expression to the identity of CLMs, and $dg^{i}$ stores
the mapping from the identity of CLMs to a set of regular expressions.


In more detail, each entry of the mapping log contains digests after
processing the request $req$ (received by the mapping log maintainer
at time $t$) on the digest stored in the previous record. Each of the
notations is explained as follows:
  \begin{itemize}
  \item $req$ can be $\add(rgx,id)$, $\del(rgx,id)$, $\newlog(cert)$,
    $\change(cert,\allowbreak\sign_{sk}(cert'),\sign_{sk'}(n,dg,t))$,
    $\blacklist(id)$, and $\finup$, respectively corresponding to a
    request to add a mapping $(rgx,id)$ of regular expression $rgx$ and
    identity $id$ of a $clog$, to delete a mapping $(rgx,id)$, to add
    a certificate $cert$ of a new $clog$, to change the certificate of
    a $clog$ from $cert$ to $cert'$, to blacklist $id$ of an existing
    $clog$, and to close the update request; where $sk$ and $sk'$ are
    signing keys associated to the certificate $cert$ and $cert'$,
    respectively; $cert$ and $cert'$ share the same subject, and $n$
    and $dg$ are the size and the digest of the corresponding $clog$
    at time $t$, respectively;

  \item $dg^s$ is the digest of an ordered data structure storing the identity
    information of the form $(cert, \sign_{sk}(n,dg,t))$ for the currently
    active certificate logs, where $cert$ is the certificate for the signing key
    $sk$ of the certificate log, and $n$ and $dg$ are respectively the size and
    digest of the certificate log at time $t$. Data are ordered by the domain
    name in $cert$.
  \item $dg^{bl}$ is the digest of an ordered data structure storing
    the domain names of blacklisted certificate logs. Data are ordered
    by the stored domain names.
  \item $dg^{r}$ is the digest of an ordered data structure storing
    elements of the form $(rgx, id)$, which represents the mapping
    from regular expression $rgx$ to the identity $id$ of a $clog$,
    data are ordered by $rgx$;
  \item $dg^{i}$ is the digest of an ordered data structure storing
    elements of the form $(id,dg^{irgx})$, which represents the
    mapping from identity $id$ of a $clog$ to a digest $dg^{irgx}$ of
    ordered data structure storing a set of regular expressions, data
    are ordered by $id$.
  \end{itemize}

\medskip

The requests are used for modifying mappings or the existing set of
certificate log maintainers. When a request $\del(rgx,id)$ has been
processed, the maintainer of certificate log with identity $id$ needs
to remove all certificates whose subject is an instance of regular
expression $rgx$; when a request $\add(rgx,id)$ has been processed,
the maintainer of certificate log with identity $id$ needs to download
all certificates whose subject is an instance of $rgx$ from the
previous authorised log maintainer, and adds them into his log. These
requests require certificate logs to synchronise with the mapping log;
see Section \ref{sec:syn}.

\subsection{Certificate logs}
The mapping log determines which certificate log is used for a domain.
The certificates for the domain are stored in that certificate log.

A certificate log mainly stores certificates for domains according to
the mappings presented in the mapping log. In particular, a
certificate log is also organised by using a chronological data
structure, and each entry of the log is of the form
  \[\hfun(req, N, dg^{rgx})\]
  where $req$ is the received request and is processed at the time
  such that the mapping log is of size $N$; $dg^{rgx}$ represents an
  ordered data structure storing a set of mappings from regular
  expressions to the information associated to the corresponding
  domains, such that the domain name is an instance of the regular
  expression. The stored information of a domain includes the identity
  and the master certificate of the domain, and two digests $dg^a$ and
  $dg^{rv}$ each presents an ordered data structure storing a set of
  active TLS certificates and a set of expired or revoked TLS
  certificates, respectively.

\medskip

\begin{figure}[!htbp]
\includegraphics[width=.5\textwidth]{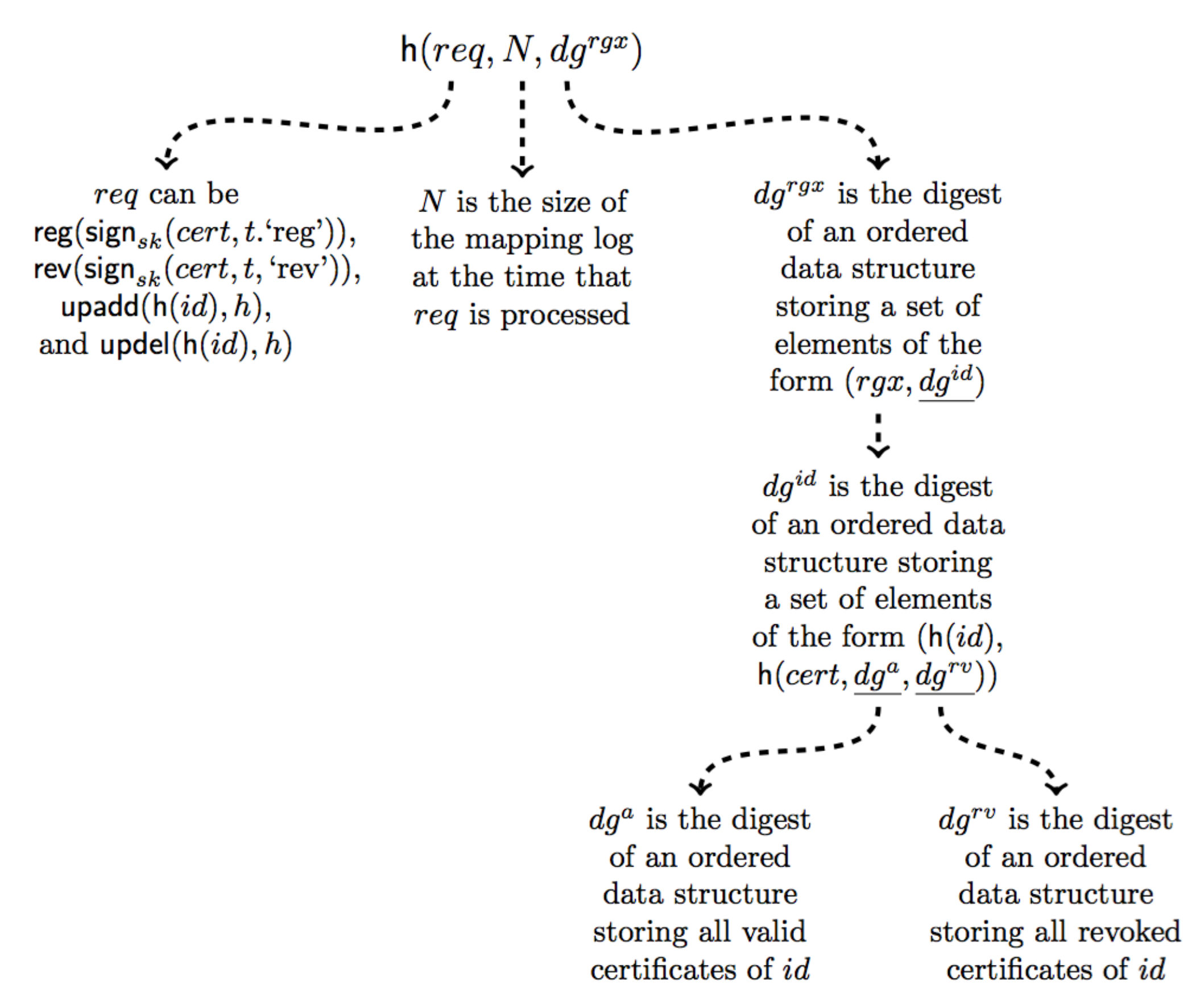}
\caption{A figure representation of the format of each record in the
  certificate log.
}
  \label{eg:clog}
\end{figure}

Elements in a record (as shown in \ref{eg:clog}) of a certificate log
are detailed as follows.
  \begin{itemize}
  \item $req$ can be $\reg(\sign_{sk}(cert,t,\text{`reg'}))$,
    $\rev(\sign_{sk}(cert,t, \text{`rev'}))$,
    $\upadd\allowbreak(\hfun(id),h)$, and $\updel(\hfun(id),h)$,
    corresponding to a request to register and revoke a certificate
    $cert$ at an agreed time $t$ such that $(cert,t,\text{`reg'})$ or
    is additionally signed by the master key $sk$, and update the
    certificate log by adding and by deleting certificates of identity
    $id$ according to the changes of $mlog$, respectively. `reg' and
    `rev' are constant, and $h$ is some value and we will explain it
    later.
  \item $N$ is the size of $mlog$ at the time $req$ is processed;

  \item $dg^{rgx}$ is the digest of an ordered data structure storing
    a set of elements of the form $(rgx, dg^{id})$, represents the
    status of the certificate log after processing the request $req$,
    and stores all the regular expressions $rgx$ that the certificate
    log is associated to. $dg^{id}$ is the digest of an ordered data
    structure storing a set of elements of the form
    $(\hfun(id),\allowbreak \hfun(cert, dg^a,dg^{rv}))$. It represents
    all domains associated to $rgx$. $id$ is an instance of $rgx$ and
    is the subject of master certificate $cert$. $dg^a$ and $dg^{rv}$
    are digests of two ordered data structures each of which
    respectively stores a set of active and revoked TLS
    certificates. In addition, data in the structure represented by
    $dg^{rgx}$ and $dg^{id}$ are ordered by $rgx$ and $h(id)$,
    respectively; data in the structure represented by $dg^a$ and
    $dg^{rv}$ are ordered by the subject of TLS certificates.
\end{itemize}

Note that requests $\upadd(\hfun(id),h)$ and $\updel(\hfun(id),h)$ are
made according to the mapping log. Even though these modifications are
not requested by domain owners, it is important to record them in the
certificate log to ensure the transparency of the log maintainer's
behaviour. Request $\upadd(\hfun(id),h)$ states that the certificate
log maintainer is authorised to manage certificates for the domain
name $id$ from now on, and the current status of certificates for $id$
is represented by $h$, where $h= \hfun(cert, dg^a,dg^{rv})$ for some
certificate $cert$ and some digest $dg^a$ and $dg^{rv}$ representing
the active and revoked certificates of $id$. $h$ is the value obtained
from the certificate log that is previously authorised to manage
certificates for domain $id$. Similarly, request $\updel(\hfun(id),h)$
indicates that the certificate log cannot manage certificates for
domain $id$ any more according to the request in the mapping log.

\subsection{Synchronising the mapping log and certificate logs}
\label{sec:syn}

The mapping log periodically (e.g. every day) publishes a signature
$\sign_{sk}(t,dg,N)$, called \emph{signed Mlog time-stamp}, on a time
$t$ indicating the publishing time, and the digest $dg$ and size $N$
of the mapping log. Mirrors of the mapping log need to download this
signed data, and update their copy of the mapping log when it is
updated. A \emph{signed Mlog time-stamp} is only valid during the issue
period (e.g. the day of issue). Note that mirrors can provide the same
set of proofs as the mapping log maintainer, because the mirror has
the copy of the entire mapping log; but mirrors are not required to be
trusted, they do not need to sign anything, and a mirror which changed
the log by itself will not be able to convince other users to accept
it since the mirror cannot forge the \emph{signed Mlog time-stamp}.

When a mapping log maintainer needs to update the mapping log, he
requests all certificate log maintainers to perform the required
update, and expects to receive the digest and size of all certificate
logs once they are updated. After the mapping log maintainer receives
these confirmations from all certificate log maintainers, he publishes
the series of update requests in the mapping log, and appends an extra
constant request $\finup$ after them in the log to indicate that the
update is done.

Log maintainers only answer requests according to their newly updated
log if the mapping log maintainer has published the update requests in
the mapping log. If in the log update period, some user sends requests
to the mapping log maintainer or certificate log maintainers, then
they give answers to the user according to their log before the update
started.

We say that the mapping log and certificate logs are
\emph{synchronised}, if certificate logs have completed the log update
according to the request in the mapping log. Note that a mis-behaving
certificate log maintainer (e.g. one recorded fake certificates in his
log, or did not correctly update his log according to the request of
the mapping log) can be terminated by the mapping log maintainer by
putting the certificate log maintainer's identity into the blacklist,
which is organised as an ordered data structure represented by
$dg^{bl}$ (as presented in \ref{sec:mlog}).

\section{Distributed transparent key infrastructure}
\label{sec: DTKI description}

Distributed transparent key infrastructure (DTKI) contains three main
phases, namely certificate publication, certificate verification, and
log verification. In the certificate publication phase, domain owners
can upload new certificates and revoke existing certificates in the
certificate log they are assigned to; in the certificate verification
phase, one can verify the validity of a certificate; and in the log
verification phase, one can verify whether a log behaves correctly.

We present DTKI using the scenario that a TLS user Alice wants to
securely communicate with a domain owner Bob who maintains the domain
$example.com$.

%

\subsection{Certificate insertion and revocation}
\label{sec:certpub}

To publish or revoke certificates in the certificate log, the domain
owner Bob needs to know which certificate log is currently authorised
to record certificates for his domain. This can be done by
communicating with a mirror of the mapping log. We detail the protocol
for requesting the mapping for Bob's domain.

\subsubsection{Request mappings}
\label{sec:reqmap}

Upon receiving the request, the mirror locates the certificate
of the authorised CLM, and generates the proofs that

\begin{enumerate}[a)]
\item the CLM is authorised for the domain; and
\item the certificate is the current valid certificate for the CLM.
\end{enumerate}

Loosely speaking, proof a) is the proof that the mapping from regular
expression $rgx$ to identity $id$ is present in the digest $dg^r$ (as
presented in the mapping log structure), such that $example.com$ is an
instance of $rgx$, and $id$ is the identity of the CLM; proof b) is
the proof that the certificate with subject $id$ is present in $dg^s$;
additionally, a proof that both $dg^s$ and $dg^r$ are present in the
latest record of the mapping log is needed. All proofs should be linked
to the latest digest signed by the MLM. If Bob has previously observed
a version of the mlog, then a proof that the current mlog is an
extension of the version that Bob observed will also be provided.

Bob accepts the response if all proofs are valid. He then stores the
verified data in his cache for future connection until the signed
digest is expired.

In more detail, after a mirror receives a request from Bob, the mirror
obtains the data of the latest element of its copy of the mapping log,
denoted $h = \hfun(req, t, dg^s, dg^{bl}, dg^r, dg^i)$, and generates
the proof of its presence in the digest (denoted $dg_{mlog}$) of its
log of size $N$. Then, it generates the proof of presence of the
element $(cert,\sign_{sk}(n,dg,t))$ in the digest $dg^s$ for some
$\sign_{sk}(n,dg,t)$, proving that the certificate log maintainer
whose $cert$ belongs to is still active. Moreover, it generates the
proof of presence of some element $(rgx,id)$ in the digest $dg^r$
where $id$ is the subject of $cert$ and $example.com$ is an instance
of the regular expression $rgx$, proving that $id$ is authorised to
store the certificates of $example.com$. The mirror then sends to Bob
the hash $h$, the signature $\sign_{sk}(n,dg,t)$, the regular
expression $rgx$, the three generated proofs of presence, and the
latest \emph{signed Mlog time-stamp} containing the time $t_{mlog}$,
and digest $dg_{mlog}$ and size $N_{mlog}$ of the mapping log.



Bob first verifies the received \emph{signed Mlog time-stamp} with the
public key of the mapping log maintainer embedded in the browser, and
verifies whether $t_{Mlog}$ is valid or not. Then Bob checks that
$example.com$ is an instance of $rgx$, and verifies the three
different proofs of presence. If all checks hold, then Bob sends the
\emph{signed Mlog time-stamp} containing
$(t'_{Mlog},dg'_{mlog},N'_{mlog})$ that he stored during a previous
connection, and expects to receive a proof of extension of
$(dg'_{mlog},N'_{mlog})$ into $(dg_{mlog},N_{mlog})$. If the received
proof of extension is valid, then Bob stores the current \emph{signed
  Mlog time-stamp}, and believes that the certificate log with
identity $id$, certificate $cert$, and size that should be no smaller
than $n$, is currently authorised for managing certificates for his
domain.

\subsubsection{Insert and revoke certificates}
\label{sec:certpub2}

The first time Bob wants to publish a certificate for his domain, he
needs to generate a pair of master signing key, denoted $sk_{m}$, and
verification key. The latter is sent to a certificate authority, which
verifies Bob's identity and issues a master certificate $cert_m$ for
Bob. After Bob receives his master certificate, he checks the
correctness of the information in the certificate. The TLS certificate
can be obtained in the same way.

To publish the master certificate, Bob signs the certificate together
with the current time $t$ by using the master signing key $sk_{m}$,
and sends it together with the request $AddReq$ to the authorised
certificate log maintainer whose signing key is denoted
$sk_{clog}$. The certificate log maintainer checks whether there
exists a valid master certificate for $example.com$; if there is one,
then the log maintainer aborts the conversation. Otherwise, the log
maintainer verifies the validity of time $t$ and the signature.

If they are all valid, the log maintainer updates the log, generates
the proof of presence that the master certificate for Bob is included
in the log, and sends the signed proof and the updated digest of the
log back to Bob. If the signature and the proof are valid, and the
size of the log is no smaller than what the mirror says, then Bob
accepts and stores the response as an evidence of successful
certificate publication. If Bob has previously observed a version of
the clog, then a proof that the current clog is an extension of the
version that Bob observed is also required.

\begin{figure}[!htbp]

\includegraphics[width=.5\textwidth]{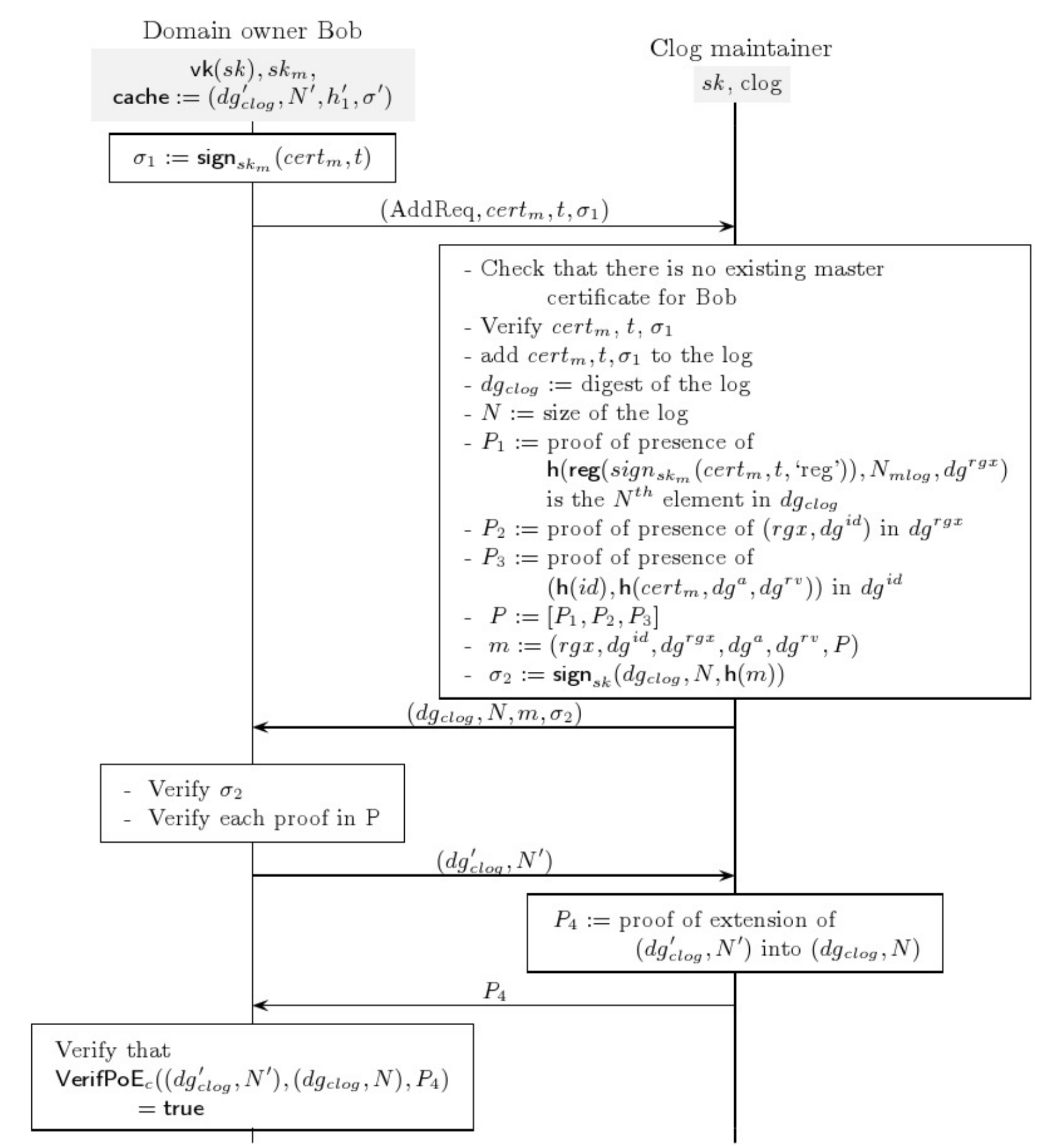}
    \caption{The protocol presenting how domain owner Bob communicates
  with certificate log (clog) maintainer to publish a master
  certificate $cert_{m}$.}
\label{fig: registration protocol}
\end{figure}

Figure~\ref{fig: registration protocol} presents the detailed process
to publish the master certificate $cert_m$. After a log maintainer
receives and verifies the request from Bob, the log maintainer updates
the log, generates the proof of presence of
$(\hfun(id),\hfun(cert_{m}, dg^a,dg^{rv}))$ in $dg^{id}$,
$(rgx, dg^{id})$ in $dg^{rgx}$, and
$\hfun(\reg(sign_{sk_m}(cert_m,t,\text{`reg'})),\allowbreak N_{mlog},
dg^{rgx})$
is the last element in the data structure represented by $dg_{clog}$,
where $id$ is the subject of $cert_m$ and an instance of $rgx$;
$\reg(sign_{sk_m}(cert_m,t, \text{`reg'}))$ is the register request to
adding $cert_m$ into the certificate log with digest $dg_{clog}$ at
time $t$. The log maintainer then issues a signature on
$(dg_{clog},N,h)$, where $N$ is the size of the certificate log, and
$h=\hfun((rgx, dg^{id}),dg^{rgx},P)$, where $P$ is the sequence of the
generated proofs, and sends the signature $\sigma_2$ together with
$(dg_{clog},N, rgx, dg^{id},dg^{rgx}, dg^a, dg^{rv}, P)$ to Bob. If
the signature and the proof are valid, and $N$ is no smaller than the
size $n$ contained in the \emph{signed Mlog time-stamp} that Bob
received from the mirror, then Bob stores the signed
$(dg_{clog}, N,h)$, sends the previous stored $(dg'_{clog},N')$ to the
certificate log maintainer, and expects to receive a proof of
extension of $(dg'_{clog},N')$ into $(dg_{clog},N)$. If the received
proof of extension is valid, then Bob believes that he has
successfully published the new certificate.

Note that it is important to send $(dg'_{clog},N')$ after receiving
$(dg_{clog},N)$, because otherwise the log maintainer could learn the
digest that Bob has, then give a pair $(dg''_{clog},N'')$ of digest
and size of the log such that $N'<N''<N$. This may open a window to
attackers who wants to convince Bob to use a certificate which was
valid in $dg''_{clog}$ but revoked in $dg_{clog}$.

In addition, if Bob has run the request mapping protocol more than
once, and has obtained a digest that is different from his local copy
of the corresponding certificate log, then he should ask the CLM to
prove that one of the digests is an extension of the other.


The process of adding a TLS certificate is similar to the process of
adding a master certificate, but the log maintainer needs to verify
that the TLS certificate is signed by the valid master signing key
corresponding to the master certificate in the log.

\medskip

To revoke a (master or TLS) certificate, the domain owner can perform
a process similar to the process of adding a new certificate. For a
revocation request with $\sign_{sk_{m}}(cert,t)$, the log maintainer
needs to check that $\sign_{sk_{m}}(cert,t')$ is already in the log
and $t>t'$. This ensures that the same master key is used for the
revocation.

\subsection{Certificate verification}
\label{sec:certverif}

When Alice wants to securely communicate with $example.com$, she sends
the connection request to Bob, and expects to receive a master
certificate $cert_{m}$ and a signed TLS certificate
$\sign_{sk_{m}}(cert,t)$ from him. To verify the received
certificates, Alice checks whether the certificates are expired. If
both of them are still in the validity time period, Alice requests (as
described in \ref{sec:reqmap}) the corresponding mapping from a mirror
to find out the authorised certificate log for $example.com$, and
communicates with the (mirror of) authorised certificate log
maintainer to verify the received certificate.

Note that this verification requests extra communication round trips,
but it gives a higher security guarantee. An alternative way is that
Bob provides both certificates and proofs, and Alice verifies the
received proofs directly.

\begin{figure}[!htbp]
\includegraphics[width=.5\textwidth]{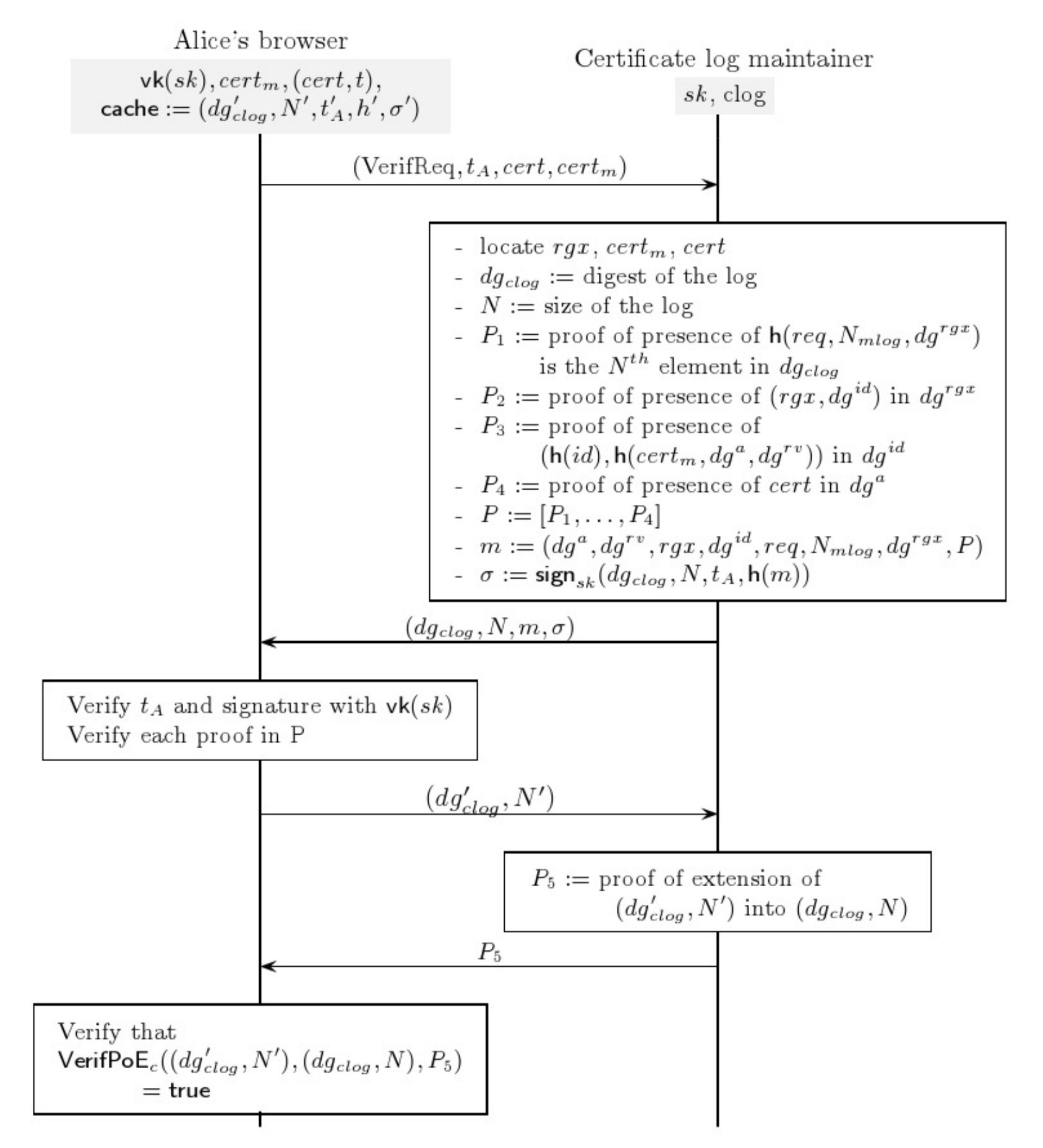}
 \caption{The protocol for verifying a certificate with the
   corresponding certificate log maintainer.}
\label{fig: request protocol}
\end{figure}

Figure \ref{fig: request protocol} presents the detailed process of
verifying a certificate. After Alice learns the identity of the
authorised certificate log, she sends the verification request
$VerifReq$ with her local time $t_A$ and the received certificate to
the certificate log maintainer. The time $t_A$ is used to prevent
replay attacks, and will later be used for accountability. The
certificate log maintainer checks whether $t_A$ is in an acceptable
time range (e.g. $t_A$ is in the same day as his local time). If it
is, then he locates the corresponding $(rgx,dg^{id})$ in $dg^{rgx}$ in
the latest record of his log such that $example.com$ is an instance of
regular expression $rgx$, locates
$(\hfun(id),\hfun(cert_{m},\allowbreak dg^a,dg^{rv}))$ in $dg^{id}$
and $cert$ in $dg^a$, then generates the proof of presence of $cert$
in $dg^a$, $(\hfun(id),\hfun(cert_{m}, dg^a,dg^{rv}))$ in $dg^{id}$,
$(rgx,dg^{id})$ in $dg^{rgx}$, and $\hfun(req,N_{mlog},dg^{rgx})$ is
the latest record in the digest $dg_{clog}$ of the log with size
$N$. Then, the certificate log maintainer signs $(dg_{clog},N,t_A,h)$,
where $h=\hfun(m)$ such that
$m= (dg^a,dg^{rv},\allowbreak rgx, dg^{id}, req, N_{mlog},
dg^{rgx},P)$,
and $P$ is the set of proofs, and sends $(dg_{clog},N,\sigma)$ to
Alice.

Alice should verify that $N_{mlog}$ is the same as her local copy of
the size of mapping log. If the received $N_{mlog}$ is greater than
the copy, then it means that the mapping log is changed (it rarely
happens) and Alice should run the request mapping protocol again. If
$N_{mlog}$ is smaller, then it means the CLM has misbehaved. Alice
then verifies the signature and proofs, and sends the previously
stored $dg'_{clog}$ with the size $N'$ to the log maintainer, and
expects to receive the proof of extension of $(dg'_{clog},N')$ into
$(dg_{clog},N)$. If they are all valid, then Alice replaces the
corresponding cache by the signed $(dg_{clog},N,t_A,h)$ and believes
that the certificate is an authentic one.

In order to preserve privacy of Alice's browsing history, instead of
asking Alice to query all proofs from the log maintainer, Alice can
send the request to Bob who will redirect the request to the log
maintainer, and redirect the received proofs from the log maintainer
to Alice.

With DTKI, Alice is able to verify whether Bob's domain has a
certificate by querying the proof of absence of certificates for
$example.com$ in the corresponding certificate log. This is useful to
prevent TLS stripping attacks, where an attacker can maliciously
convert an HTTPS connection into an HTTP connection.

\subsection{Log verification}
\label{sec:logverif}

Users of the system need to verify that the mapping log maintainer and
certificate log maintainers did update their log correctly according
to the requests they have received, and certificate log maintainers
did follow the latest mappings specified in the mapping log.

These checks can be easily done by a trusted monitor.  However, since
we aim to provide a TTP-free system, DTKI uses a crowdsourcing-like
method, based on random checking, to monitor the correctness of the
public log. The basic idea of random checking is that each user
randomly selects a record in the log, and verifies whether the request
and data in this record have been correctly managed. If all records
are verified, the entire log is verified. Users only need to run the
random checking periodically (e.g. once a day). The full version (with
formalisation) of random checking can be found in our technical
report. We give a flavour here by providing an example. Example
\ref{ex:mlog} presents the random checking process to verify the
correct behaviour of the mapping log.

\begin{example}
\label{ex:mlog}
Suppose verifier has randomly selected the $k^{\text{th}}$ record of
the mapping log, and the record has the form
$\hfun(\add(rgx,id),t_k,dg^s_k,dg^{bl}_k,dg^r_k,dg^i_k)$. The verifier
must check that all digests in this record are updated from the
$(k-1)^{\text{th}}$ record by adding a new mapping $(rgx,id)$ in the
mapping log at time $t_k$.

Let the label of the $(k-1)^{\text{th}}$ record be
$\hfun(req_{k-1},t_{k-1},dg^s_{k-1}, \allowbreak
dg^{bl}_{k-1},dg^r_{k-1},dg^i_{k-1})$,
then to verify the correctness of this record, the verifier should run
the following process:

\begin{itemize}
\item verify that $dg^s_{k}=dg^s_{k-1}$ and
  $dg^{bl}_{k}=dg^{bl}_{k-1}$; and
\item verify that $dg^r_k$ is the result of adding $(rgx,id)$ into
  $dg^r_{k-1}$ by using $\OLVerifPoAddName$, and $id$ is an instance
  of $rgx$; and

\item verify that $(id,dg_k^{irgx})$ is the result of replacing
  $(id,dg_{k-1}^{irgx})$ in $dg_{k-1}^i$ by $(id,dg_k^{irgx})$ by
  using $\OLVerifPoMName$; and

\item verify that $dg_k^{irgx}$ is the result of adding $rgx$ into
  $dg_{k-1}^{irgx}$ by using $\OLVerifPoAddName$.
\end{itemize}
Note that all proofs required in the above are given by the log
maintainer. If the above tests succeed, then the mapping log
maintainer has behaved correctly for this record.
\end{example}

\subsection{Performance Evaluation}
\label{sec:p}

In this section, we measure the cost of different protocols in DTKI.

\paragraph*{{\bf Assumptions}}
We assume that the size of a certificate log is $10^8$ (the total
number of registered domain names currently is $2.71\times 10^8$
\cite{NumofDNS}, though only a fraction of them have certificates). In
addition, we assume that the number of stored regular expressions, the
number of certificate logs, and the size of the mapping log are 1000
each. (In fact, if we assume a different number or size (e.g. 100 or
10000) for them, it makes almost no difference to the conclusion).
Moreover, in the certificate log, we assume that the size of the set
of data represented by $dg^{rgx}$ is 10, by $dg^{id}$ is $10^5$, by
$dg^a$ is 10, and by $dg^{rv}$ is 100. These assumptions are based on
the fact that $dg^{rgx}$ represents the set of regular expressions
maintained by a certificate log; the $dg^{id}$ represents the set of
domains which is an instance of a regular expression; and $dg^a$ and
$dg^{rv}$ represent the set of currently valid certificates and the
revoked certificates, respectively. Furthermore, we assume that the
size of a certificate is 1.5 KB, the size of a signature is 256 bytes,
the length of a regular expression and an identity is 20 bytes each,
and the size of a digest is 32 bytes.

\paragraph*{{\bf Space}}
Based on these assumptions, the approximate size of the transmitted
data in the protocol for publishing a certificate is 4 KB, for
requesting a mapping is 3 KB, and for verifying a certificate is 5
KB. Since the protocols for publishing a certificate and requesting a
mapping are run occasionally, we mainly focus on the cost of the
protocol for verifying a certificate, which is required to be run
between a log server and a web browser in each secure connection.

By using Wireshark, we\footnote{We use a MacBook Air 1.8 GHz Intel
  Core i5, 8 GB 1600 MHz DDR3.} measure that the size of data for
establishing an HTTPS protocol to log-in to the internet bank of HSBC,
Bank of America, and Citibank are 647.1 KB, 419.9 KB, and 697.5 KB,
respectively. If we consider the average size ($\approx$588 KB) of
data for these three HTTPS connections, and the average size
($\approx$6 KB) of data for their corresponding TLS establishment
connections, we have that in each connection, DTKI incurs 83$\%$
overhead on the cost of the TLS protocol. However, since the total
overhead of an HTTPS connection is around 588 KB, so the cost of DTKI
only adds 0.9$\%$ overhead to each HTTPS connection, which we consider
acceptable.

\paragraph*{{\bf Time}}
Our implementation uses a SHA-256 hash value as the digest of a log
and a 2048 bit RSA signature scheme. The time to compute a
hash\footnote{SHA-256 on 64 byte size block.} is $\approx 0.01$
millisecond (ms) per 1KB of input, and the time to verify a 2048 bit
RSA signature is 0.48 ms. The approximate verification time on the
user side needed in the protocol for verifying certificates is 0.5 ms.

\medskip

Hence, on the user side, the computational cost on the protocol for
verifying certificates incurs 83$\%$ on the size of data for
establishing a TLS protocol, and 0.9$\%$ on the size of data for
establishing an HTTPS protocol; the verification time on the protocol
for verifying certificates is 1.25 $\%$ of the time for establishing a
TLS session (which is approximately 40 ms measured with Wireshark on
the TLS connection to HSBC bank).

\section{Security Analysis}
\label{sec:security}

We consider an adversary who can compromise the private key of all
infrastructure servers in DTKI. In other words, the adversary can
collude with all log servers and certificate authorities to launch
attacks.

\paragraph*{\textbf{Main result}}
Our security analysis shows that
\begin{itemize}
\item if the distributed random checking has verified all required
  tests, and domain owners have successfully verified their initial
  master certificates, then DTKI can prevent attacks from the
  adversary; and

\item if the distributed random checking has not completed all
  required tests, or domain owners have not successfully verified
  their initial master certificates, then an adversary can launch
  attacks, but the attacks will be detected afterwards.
\end{itemize}

We provide all source codes and files required to understand and
reproduce our security analysis at~\cite{tamarin-DTKI}. In particular,
these include the complete DTKI models and the verified proofs.

\subsection{Formal analysis}

We analyse the main security properties of the DTKI protocol using the
\Tamarin~prover~\cite{Tamarin}. The \Tamarin~prover is a symbolic
analysis tool that can prove properties of security protocols for an
unbounded number of instances and supports
reasoning about protocols with mutable global state, which makes it
suitable for our log-based protocol. Protocols are specified using multiset
rewriting rules, and properties are expressed in a guarded fragment of
first order logic that allows quantification over timepoints.

\Tamarin is capable of automatic verification in many cases, and it also
supports interactive verification by manual traversal of the proof tree. If the
tool terminates without finding a proof, it returns a counter-example.
Counter-examples are given as so-called dependency graphs, which are partially
ordered sets of rule instances that represent a set of executions that
violate the property. Counter-examples can be used to refine the model, and
give feedback to the implementer and designer.

\paragraph*{\textbf{Modeling aspects}}
We used several abstractions during modeling. We model our log as
lists, similar to the abstraction used in~\cite{ARPKI,KUD}. We also
assume that the random checking is verified.

We model the protocol roles D (domain server), M (mapping log
maintainer), C (certificate log maintainer), and CA (certificate
authority) by a set of rewrite rules. Each rewrite rule typically
models receiving a message, taking an appropriate action, and sending
a response message.  Our modeling approach is similar to the one used
in most \Tamarin models. Our modeling of the roles directly
corresponds to the protocol descriptions in the previous sections.
\Tamarin provides built-in support for a Dolev-Yao style network
attacker, i.e., one who is in full control of the network. We
additionally specify rules that enable the attacker to compromise
service providers, namely the mapping log maintainer, certificate log
maintainers and CAs, learn their secrets, and modify public logs.

Our final DTKI model (available from \cite{tamarin-DTKI}) consists of
959 lines for the base model and five main property specifications,
examples of which we will give below.

\paragraph*{\textbf{Proof goals}}
We state several proof goals for our model, exactly as specified in
\Tamarin's syntax. Since \Tamarin's property specification language is a
fragment of first-order logic, it contains logical connectives ({\tt |},
{\tt \&}, {\tt ==>}, {\tt not}, ...) and quantifiers ({\tt All}, {\tt Ex}).
In \Tamarin, proof goals are marked as
{\tt lemma}. The {\tt \#}-prefix is used to denote timepoints, and ``{\tt E @
\#i}'' expresses that the event $E$ occurs at timepoint $i$.

The first goal is a check for executability that ensures that our model
allows for the successful transmission of a message. It is encoded in the
following way.

{\scriptsize
\begin{verbatim}
lemma protocol_correctness:
 exists-trace
 " /* It is possible that */
   Ex D Did m rgx ltpkD stpkD #i1.

   /* The user has sent an encrypted message aenc{m}stpkD
      to domain server D whose identity is Did and TLS key
      is stpk, and the user received from D a confirmation
      h(m) of receipt. */

       Com_Done(D, Did, m, rgx, ltpkD, stpkD) @ #i1

     /* without the adversary compromising any party. */
   &   not (Ex #i2 CA ltkCA.
                Compromise_CA(CA,ltkCA) @ #i2)

   &   not (Ex #i3 C ltkC.
                Compromise_CLM(C,ltkC) @ #i3)

   &   not (Ex #i4 M ltkM.
                Compromise_MLM(M,ltkM) @ #i4)

"
\end{verbatim}
}

The property holds if the \Tamarin model exhibits a behaviour in which
a domain server received a message without the attacker compromising
any service providers.  This property mainly serves as a sanity check
on the model.  If it did not hold, it would mean our model does not
model the normal (honest) message flow, which could indicate a flaw in
the model.  \Tamarin automatically proves this property in several
minutes and generates the expected trace in the form of a graphical
representation of the rule instantiations and the message flow.

We additionally proved several other sanity-checking properties to
minimize the risk of modeling errors.

The second example goal is a secrecy property with respect to a
classical attacker, and expresses that when no service provider is
compromised, the attacker cannot learn the message exchanged between a
user and a domain server. Note that {\tt K(m)} is a special event that
denotes that the attacker knows $m$ at this time.

{\scriptsize
\begin{verbatim}
lemma message_secrecy_no_compromised_party:
 "
 All D Did m rgx ltpkD stpkD #i1.

   /* The user has sent an encrypted message aenc{m}stpkD
      to domain server D whose identity is Did and TLS key
      is stpk, and the user received from D a confirmation
      h(m) of receipt. */

       (Com_Done(D, Did, m, rgx, ltpkD, stpkD) @ #i1

/* and no party has been compromised */
   &   not (Ex #i2 CA ltkCA.
                Compromise_CA(CA,ltkCA) @ #i2)

   &   not (Ex #i3 C ltkC.
                Compromise_CLM(C,ltkC) @ #i3)

   &   not (Ex #i4 M ltkM.
                Compromise_MLM(M,ltkM) @ #i4)
      )
      ==>
      ( /* then the adversary cannot know m */
       not (Ex #i5. K(m) @ #i5)
      )
 "
\end{verbatim}
}

\Tamarin proves this property automatically (in 575 steps).

The above result implies that if a domain server D, whose domain name
is Did such that Did is an instance of regular expression rgx,
receives a message that was sent by a user, and the attacker did not
compromise server providers, then the attacker will not learn the
message.

The next two properties encode the unique security guarantees provided
by our protocol, in the case that even all service providers are
compromised.

The first main property we prove is that when all service providers
(i.e. CAs, the MLM, and CLMs) are compromised, and the domain owner
has successfully verified his master certificate in the log, then the
attacker cannot learn the message exchanged between a user and a
domain owner. It is proven automatically by \Tamarin in 5369 steps.

{\scriptsize
\begin{verbatim}
lemma message_secrecy_compromise_all_domain_verified_master_cert:
 "
 All D Did m rgx ltpkD stpkD #i1.

   /* The user has sent an encrypted message aenc{m}stpkD to domain
   server D whose identity is Did and TLS key is stpk, and the user
   received from D a confirmation h(m) of receipt. */

   (Com_Done(D, Did, m, rgx, ltpkD, stpkD) @ #i1

   /* and at an earlier time, the domain server has verified his
      master certificate */
     & Ex #i2.
     VerifiedMasterCert(D, Did, rgx, ltpkD) @ #i2
     & #i2 < #i1

      )
      ==>
      ( /* then the adversary cannot know m */
       not (Ex #i3. K(m) @ #i3)
      )
 "
\end{verbatim}
}

The property states that if a domain server D receives a message that
was sent by a user, and at an earlier time, the domain server has
verified his master certificate, then even if the attacker can
compromise all server providers, the attacker cannot learn the
message.

The final property states that when all service providers can be
compromised, and a domain owner has not verified his/her master
certificate, and the attacker learns the message exchanged between a
user and the domain owner, then afterwards the domain owner can detect
this attack by checking the log. It is also verified by \Tamarin
within a few minutes.

{\scriptsize
\begin{verbatim}
lemma detect_bad_records_in_the_log_when_master_cert_not_verified:

 "
 All D Did m rgx ltpkD flag stpkD #i1 #i2 #i3.

   /* The user has sent an encrypted message aenc{m}stpkD to domain
   server D whose identity is Did and TLS key is stpk, and the user
   received from D a confirmation h(m) of receipt. */

   (Com_Done(D, Did, m, rgx, ltpkD, stpkD) @ #i1

   /* and the adversary knows m */

     &  K(m) @ #i2

   /* and we afterwards check the log */
     &  CheckedLog(D, Did, rgx, ltpkD, flag, stpkD) @ #i3
     &   #i1 < #i3)
      ==>
     ( /* then we can detect a fake record in the log */
        (flag = 'bad')
     )
 "
\end{verbatim}
}
\section{Comparison}
\label{sec:comparison}
\begin{figure*}[htp]
\center
 \small
\resizebox{0.92\textwidth}{!}{
\begin{tabular}{>{\raggedright}p{.30\textwidth}|
                >{\raggedright}p{.1\textwidth}|
                >{\raggedright}p{.1\textwidth}|
                >{\raggedright}p{.1\textwidth}|
                >{\raggedright}p{.1\textwidth}|
 >{\raggedright\arraybackslash}p{.15\textwidth}
 @{}}

  & \multicolumn1{c|}{SK \cite{SovereignKeys}}
  & \multicolumn1{c|}{CT \cite{rfc6962}}
  & \multicolumn1{c|}{AKI \cite{AKI}}
  & \multicolumn1{c|}{ARPKI \cite{ARPKI}}
  & \multicolumn1{c}{DTKI}\\\hline
  \multicolumn1{@{}l|}{\bf Terminology}&&&&&\\[.5ex]
  Log provider
  & Time-line server
  & Log
  & Integrity log server (ILS)
  & Integrity log server (ILS)
  & Certificate/Mapping log maintainer (CLM, MLM)\\[1ex]
  Log extension
  & \multicolumn1{c|}{-}
  & Log consistency
  & \multicolumn1{c|}{-}
  & \multicolumn1{c|}{-}
  & Log extension\\[1ex]
  Trusted party
  & Mirror
  & Auditor \& monitor
  & Validator
  & Validator (optional)
  & \multicolumn1{c}{-}\\[2ex]\hline
  \multicolumn1{@{}p{.3\textwidth}|}{\bf Whether answers to queries rely on trusted parties or are accompanied by a proof}&&&&&\\[6ex]
  Certificate-in-log query:
  &  Rely
  &  Proof
  &  Proof
  &  Proof
  &  Proof \\
  Certificate-current-in-log query:
  &  Rely
  &  Rely
  &  Proof
  &  Proof
  &  Proof \\
  Subject-absent-from-log query:
  &  Rely
  &  Rely
  &  Proof
  &  Proof
  &  Proof \\
  Log extension query:
  &  Rely
  &  Proof
  &  Rely
  &  Rely
  &  Proof \\[2ex]\hline
  \multicolumn1{@{}l|}{\bf Non-necessity of trusted monitors}&&&&&\\[.5ex]
  The role of trusted monitors can be distributed to browsers
  & No
  & No
  & No$^{+}$
  & No$^{+}$
  & Yes\\[4ex]\hline
  \multicolumn1{@{}l|}{\bf Trust assumptions}&&&&&\\[.5ex]
  Not all service providers collude together
  & Yes
  & Yes
  & Yes
  & Yes
  & No\\ &&&&\\[-1.5ex]
  Domain is initially registered by an honest party
  & No
  & No
  & Yes*
  & Yes*
  & Yes*\\[4ex]\hline
  \multicolumn1{@{}l|}{\bf Security guarantee}&&&&&\\[.5ex]
  Attacks detection or prevention
  & Detection
  & Detection
  & Prevention
  & Prevention
  & Prevention\\[2ex]\hline
  \multicolumn1{@{}l|}{\bf Oligopoly issues}&&&&&\\[.5ex]
  Log providers required to be built into browser (oligopoly)
  & Yes
  & Yes
  & Yes
  & Yes
  & Only MLM\\ &&&&\\[-1.5ex]
  Monitors required to be built into browser (oligopoly and trust non-agility)
  & Yes
  & No
  & Yes
  & Yes$^{\dagger}$
  & No\\
  \hline
  \multicolumn6{@{}l}{}\\[.5ex]
  \multicolumn6{@{}l}{+ The system limits the trust in each server by letting them to monitor each other's behaviour.}\\[.5ex]
  \multicolumn6{@{}l}{* Without the assumption, the security guarantee is detection rather than prevention.}\\[.5ex]
  \multicolumn6{@{}l}{$\dagger$ The trusted party is optional, if there is a trusted party, then the trusted party is required to be built into browser.}\\[0.5ex]
  \end{tabular}
}
  \caption{Comparison of log-based approaches to certificate
    management. {\bf Terminology} helps compare the terminology used
    in the papers. {\bf How queries rely on trusted parties} shows
    whether responses to browser queries come with proof of
    correctness or rely on the honesty of trusted parties. {\bf
      Necessity of trusted parties} shows whether the TP role can be
    performed by browsers. {\bf Trust assumptions} shows the
      assumption for the claimed security guarantee. {\bf Oligopoly
      issues} shows the entities that browsers need to know about.}
  \label{fig:comparison}
\end{figure*}

As mentioned previously, DTKI builds upon a wealth of ideas from SK
\cite{SovereignKeys}, CT \cite{rfc6962}, CIRT \cite{RyanNDSS14}, and
AKI \cite{AKI}. Figure \ref{fig:comparison} shows the dimensions along
which DTKI aims to improve on those systems.

Compared with CT, DTKI supports revocation by enabling log providers
to offer proofs of \emph{absence} and \emph{currency} of certificates.
In CT, there is no mechanism for revocation.  CT has proposed
additional data structures to hold revoked certificates, and those
data structures support proofs of their contents
\cite{revocation-transparency}. However, there is no mechanism to
ensure that the data structures are maintained correctly in time.

Compared to CIRT, DTKI extends the log structure of CIRT to make it
suitable for multiple log maintainers, and provides a stronger
security guarantee as it prevents attacks rather than merely detecting
them. In addition, the presence of the mapping log maintainer and
multiple certificate log maintainers create some extra monitoring
work. DTKI solves it by using a detailed crowd-sourcing verification
system to distribute the monitoring work to all users' browsers.

Compared to AKI and ARPKI, in DTKI the log providers can give proof
that the log is maintained append-only from one step to the next. The
data structure in A(RP)KI does not allow this, and therefore they
cannot give a verifiable guarantee to the clients that no data is
removed from the log. 

DTKI improves the support that CT and A(RP)KI have for multiple log
providers.  In CT and AKI, domain owners wishing to check if there
exists a log provider that has registered a certificate for him has to
check all the log providers, and therefore the full set of log
providers has to be fixed and well-known. This prevents new log
providers being flexibly created, creating an oligopoly. In contrast,
DTKI requires the browsers only to have the MLM public key built-in,
minimising the oligopoly element.


In DTKI, trusted monitors are optional, as it uses crowd-sourced
verification. More precisely, a trusted monitor's verification work
can be done probabilistically in small pieces by users' browsers.

Unlike the mentioned previous work, DTKI allows the possibility that
all service providers (i.e. the MLM, CLMs, and mirrors) to collude
together, and can still prevent attacks. In contrast, SK and CT can
only detect attacks, and to prevent attacks, A(RP)KI requires that not
all service providers collude together. Similar to A(RP)KI, DTKI also
assumes that the domain is initially registered by an honest party to
prevent attacks, otherwise A(RP)KI and DTKI can only detect attacks.
\section{Discussion}
\label{sec:discussion}

\paragraph*{{\bf Responding to incorrect proofs}}

How should the browser (and the user) respond if a received proof
(e.g., a proof of presence in the log) is incorrect?  Such situations
should be handled in the background by the software in the browser
that verifies proofs, and be sent to domain owners for further
investigation. The browser can also present errors to the user in the
same way as the current state of the art. So, the user interface will
remain the same. For example, a user might be shown two options, i.e.
either to continue anyway, or not to trust the certificate and abort
this connection. Another possible way is to hard fail if the
verification has not been successful, as suggested by Google
certificate transparency. However, this might be an obstacle for
deploying DTKI in early stages.

\paragraph*{{\bf Coverage of random checking}}
As mentioned previously, several aspects of the logs are verified by
user's browsers performing randomly-chosen checks. The number of
things to be checked depends on the size of the mapping log and
certificate logs. The size of the mapping log mainly depends on the
number of certificate logs and the mapping from regular expressions to
certificate logs; and the size of certificate logs mainly depends on
the number of domain servers that have a TLS certificate. Currently,
there are $2.71\times 10^8$ domains \cite{NumofDNS} (though not every
domain has a certificate), and $3\times 10^9$ internet users
\cite{InternetUser}. The probability of a given domain not being
checked on a given day (or week) is
$(1-\frac{1}{2.71\times 10^{8}})^{3\times 10^{9}} \approx 1.56\times
10^{-5}$
(resp.
$((1-\frac{1}{2.71\times 10^{8}})^{3\times 10^{9}})^7 \approx
2.25\times 10^{-34}$).
Thus, the expected number of unchecked domains per day (resp. per
week) is $4.23\times 10^3$ (resp. $6.10 \times 10^{-26}$).


\paragraph*{\bf Accountability of mis-behaving parties}
The main goal of new certificate management schemes such as CT, CIRT,
AKI, ARPKI and DTKI is to address the problem of mis-issued
certificates, and to make the mis-behaving (trusted) parties
accountable.

In DTKI, a domain owner can readily check for rogue certificates for
his domain. First, he queries a mirror of the mapping log maintainer
to find which certificate log maintainers (CLM) are allowed to log
certificates for the domain (section \ref{sec: DTKI
  description}). Then he examines the certificates for his domain that
have been recorded by those CLMs. The responses he obtains from the
mirror and the CLMs are accompanied by proofs. If he detects a
mis-issued certificate, he requests revocation in the CLM. If that is
refused, he can complain to the top-level domain, who in turn can
request the MLM to change the CLM for his domain (after that, the
offending CLM will no longer be consulted by browsers). This request
should not be refused because the MLM is governed by an international
panel. The intervening step, of complaining to the top-level domain,
reflects the way domain names are actually managed in
practice. Different top-level domains have different terms and
conditions, and domain owners take them into account when purchasing
domain names. In DTKI, log maintainers are held accountable because
they sign and time-stamp their outputs. If a certificate log
maintainer issues an inconsistent digest, this fact will be detected
and the log maintainer can be blamed and blacklisted. If the mapping
log misbehaved, then its governing panel must meet and resolve the
situation.

In certificate transparency, this process is not as smooth. Firstly,
the domain owner doesn't get proof that the list of issued
certificates is complete; he needs to rely on monitors and auditors.
Next, the process for raising complaints with log maintainers who
refuse revocation requests is less clear (indeed, the RFC
\cite{rfc6962} says that the question of what domain owners should do
if they see an incorrect log entry is beyond scope of their
document). In CT, a domain owner has no ability to dissociate himself
from a log maintainer and use a different one.

AKI addresses this problem by saying that a log maintainer that refuses
to unregister an entry will eventually lose credibility through a
process managed by validators, and will be subsequently ignored. The
details of this credibility management are not very clear, but it does
not seem to offer an easy way for domain owners to control which log
maintainers are relied on for their domain.
\paragraph*{{\bf Master certificate concerns}}
One concern is that a CA might publish fake master certificates for domains that
the CA doesn't own and are not yet registered. However, this problem is not
likely to occur: CAs are businesses, they cannot afford the bad press from
negative public opinion and they cannot afford the loss of reputation. Hence,
they will only want to launch attacks that would not be caught. (Such an
adversary model has been described by Franklin and Yung \cite{FranklinY92},
Canetti and Ostrovsky \cite{CanettiO99}, Hazay and Lindell \cite{covert08}, and
Ryan \cite{RyanNDSS14}). In DTKI, if a CA attempts to publish a fake
master certificate for some domain, it will have to leave evidence of its
misbehaviour in the log, and the misbehaviour will eventually be detected by the
genuine domain owner.

Another concern is the assumption that the domain owners can securely
handle their master keys. In practice, the domain owners might have
problems looking after their master keys due to lack of awareness of
good practices. This problem arises in any web PKI: it is assumed that
domain owners can securely handle their TLS keys. Our system adds one
more key (the master key) to that requirement.  A possible practical
solution for domain owners is to use a trustworthy service to handle
TLS keys (and the master key); the details are beyond the scope of the
paper.

\paragraph*{{\bf Avoidance of oligopoly}}
As we mentioned in the introduction, the predecessors (SK, CT, CIRT,
AKI, ARPKI) of DTKI do not solve a foundational issue, namely
\emph{oligopoly}. These proposals require that all browser vendors
agree on a fixed list of log maintainers and/or validators, and build
it into their browsers. This means there will be a large barrier to
create a new log maintainer.

CT has some support for multiple logs, but it doesn't have any method
to allocate different domains to different logs. In CT, when a domain
owner wants to check whether mis-issued certificates are recorded in
logs, he needs to contact all existing logs, and download all
certificates in each of the logs, because there is no way to prove to
the domain owner that no certificates for his domain is in the log, or
to prove that the log maintainer has showed all certificates in the
log for his domain to him. Thus, to be able to detect fake
certificates, CT has to keep a very small number of log
maintainers. This prevents new log providers being flexibly created,
creating an oligopoly.

\medskip

In contrast to its predecessors, DTKI does not have a fixed set of
certificate log maintainers (CLMs) to manage certificates for domain
owners, and it allows operations of adding or removing a certificate
log maintainer by updating the mapping log. In DTKI, the public log of
the MLM is the only thing that browsers need to know.

The MLM may be thought to represent a monopoly; to the extent that it
does, it is likely to be a much weaker monopoly than the oligopoly of
CAs or log maintainers. CAs and log maintainers offer commercial
services and compete with each other, by offering different levels of
service at different price points in different markets. The MLM should
not offer any commercial services; it should perform a purely
administrative role, and is not required to be trusted because it
behaves fully transparently and does not manage any certificates for
web domains. In addition, the MLM is expected to be operated by an
international panel with a lot of members.

In practice, we expect ICANN to be the MLM, as it is responsible for
coordinating name-spaces of the Internet, and is governed by a
Governmental Advisory Committee containing representatives from 111
states. However, there might be concerns here, including the concern
that ICANN might not be interested in being the MLM, due to the fact that
the service won't generate any revenue. Our solution does not address
political issues around making decisions of whether to add or remove
some CLMs or not.






\section{Conclusions and future work}

Sovereign keys (SK), certificate transparency (CT), accountable key
infrastructure (AKI), certificate issuance and revocation transparency
(CIRT), and attack resilient PKI (ARPKI) are recent proposals to make
public key certificate authorities more transparent and verifiable, by
using public logs. CT is currently being implemented in servers and
browsers. Google is building a certificate transparency log containing
all the current known certificates, and is integrating verification of
proofs from the log into the Chrome web browser.

Unfortunately, as it currently stands, CT risks creating an oligopoly
of log maintainers (as discussed in section \ref{sec:discussion}), of
which Google itself will be a principal one. Therefore, adoption of CT
risks investing more power about the way the internet is run in a
company that arguably already has too much power.

In this paper we proposed DTKI -- a transparent public key validation
system using an improved construction of public logs. DTKI can prevent
attacks based on mis-issued certificates, and minimises undesirable
oligopoly situations by using the mapping log. In addition, we
formalised the public log structure and its implementation; such
formalisation work was missing in the previous systems (i.e. SK, CT,
A(RP)KI, and CIRT). Since devising new security protocols is
notoriously error-prone, we provide a formalisation of DTKI, and
formally proved its security properties by using Tamarin prover.

\section*{Acknowledgments}
This work was supported in part by The University of Birmingham
under EPSRC Grant EP/H005501/1. We thank anonymous reviewers for their
feedback.

\appendix

\section*{Implementation of data structures}
\label{sec:DSimplementation}
This section shows the implementation of the chronological data
structure and ordered data structure. We give some examples to show
how the proofs could be done. Full details can be found in our
technical report. We consider a secure hash function (e.g. SHA256),
denoted $\hfun$.

\subsection*{Chronological data structure}
The chronological data structure is implemented based on Merkle tree
structure that we call \emph{ChronTree}.

A \emph{ChronTree} $T$ is a binary tree whose nodes are labelled by
bitstrings such that:
\begin{itemize}
\item every non-leaf node in $T$ has two children, and is labelled
  with $\hfun(t_\ell, t_r)$ where $t_\ell$ (resp. $t_r$) is the label
  of its left child (resp. right child); and
\item the subtree rooted by the left child of a node is perfect, and
 its height is greater than or equal to the height of the subtree
 rooted by the right child.
\end{itemize}
Here, a subtree is ``perfect'' if its every non-leaf node has two
children and all its leaves have the same depth.

\begin{figure}[!htbp]
\includegraphics[width=.5\textwidth]{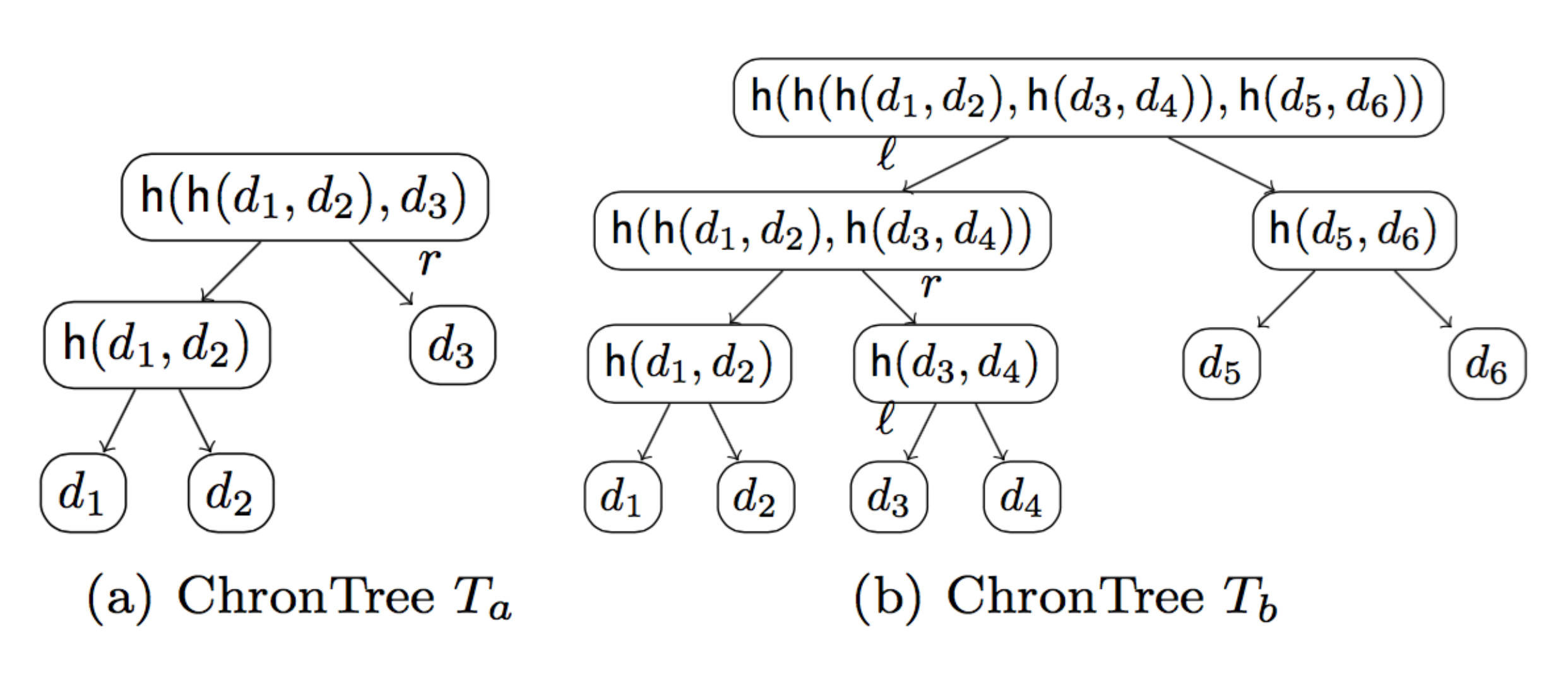}
\caption{Example of two ChronTrees, $T_a$ and $T_b$.}
\label{fig:example verification proof of extension CT}
\label{fig:example ChronTree}
\label{fig:example proof of presence CT}
\end{figure}

Note that a ChronTree is a not necessarily a balanced tree. The two
trees in Figure~\ref{fig:example ChronTree} are examples of ChronTrees
where the data stored are the bitstrings denoted $d_1, \ldots, d_6$.

\begin{figure}[!htbp]
\includegraphics[width=.5\textwidth]{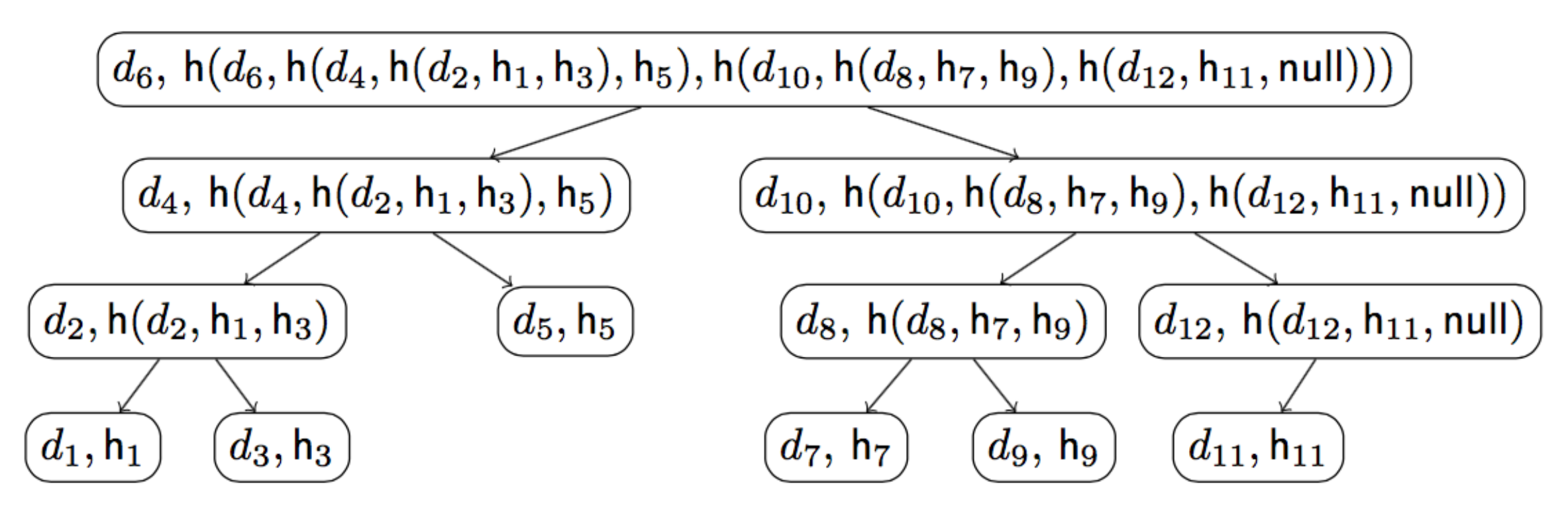}
\caption{An example of a LexTree $T_c$, where
  $\hfun{_i}=\hfun(d_i,\Null,\Null)$ for all $i=\{1,3,5,7,9,11\}$}
\label{fig:example LexTree}
\end{figure}

Given a ChronTree $T$ with $k$ leaves, we denote by $\Seq{T} =
[d_1,\ldots, d_k]$ the sequence of bitstrings stored in $T$.  Note
that a ChronTree is completely defined by the sequence of data stored
in the leaves.  Moreover, we say that the size of a ChronTree is the
number its leaves.

\medskip

Given a bitstring $d$ and a ChronTree $T$, the \emph{proof of presence
  of $d$ in $T$} exists if there is a leaf $n_1$ in $T$ labelled by
$d$; and is defined as $\CTPoP{w}{[b_1, \ldots, b_k]}$ such that:
\begin{itemize}
\item $w$ is the position in $\{\ell,r\}^*$ of $n_1$ (that is, the
  sequence of left or right choices which lead from the root to
  $n_1$), and $|w| = k$; and
\item if $n_1, \ldots, n_{k+1}$ is the path from $n_1$ to the root,
  then for all $i \in \{1, \ldots, k\}$, $b_i$ is the label of the
  sibling node of $n_i$.
\end{itemize}

Intuitively, a proof of presence of $d$ in $T$ contains the minimum
amount of information necessary to recompute the label of the root of
$T$ from the leaf containing $d$.

\begin{example}
\label{ex: proof of presence CT}
Consider the ChronTree $T_b$ of Figure~\ref{fig:example
  ChronTree}. The proof of presence of $d_3$ in $T_b$ is the tuple
$\CTPoP{w}{seq}$ where:
\begin{itemize}
\item $w = \ell\cdot r \cdot \ell$
\item $seq = [ d_4, \hfun(d_1,d_2), \hfun(d_5,d_6) ]$
\end{itemize}
\end{example}

Note that the size of the proof of presence is logarithmic in the size
of the tree; even if the tree grows considerably, the size of the
proof does not increase much.

\medskip

Let $T$ and $T'$ be ChronTrees of size $N$ and $N'$, respectively,
such that $N' \leq N$,
$\Seq{T} = [ d_1, \ldots, d_{N'}, \ldots, d_N ]$, and
$\Seq{T'} = [ d_1, \ldots, d_{N'} ]$ for some bitstrings $d_1$,
$\ldots$, $d_{N'}$, $\ldots$, $d_N$. Let $m$ be the smallest position
of the bit $1$ in the binary representation of $N'$; and let $(d,w)$
be the $(m+1)^{\text{th}}$ node in the path of the node labelled by
$d_{N'}$ to the root in $T$, where $d$ is a bitstring and
$w\in \{\ell,r\}^*$ indicates the position. At last, let $(w,seq')$ be
the proof of presence of $d$ in $T$. The proof of extension of $T'$
into $T$ is defined as the sequence $seq$ of bitstrings such that
\begin{itemize}
\item if $N' = 2^{k}$ for some $k$, then $seq=seq'$; otherwise
\item $seq=d :: seq'$, where $::$ is the concatenation operation.
\end{itemize}

\begin{example}
\label{ex: proof of extension CT}
The proof of extensions of $T_a$ into $T_b$ (Figure~\ref{fig:example
  ChronTree}) is the sequence $seq = [d_3, d_4, \hfun(d_1,d_2),
\hfun(d_5,d_6)]$.
\end{example}

While a proof of presence is the minimal amount of information
necessary to recompute the hash value of a ChronTree from the leaf
containing some particular data, the proof of extension is the minimal
amount of information necessary to recompute the hash value of
ChronTree $T$ from the hash value of a ChronTree $T'$ where $T$ is an
extension of $T'$. Intuitively, the proof of extension of a ChronTree
$T'$ into a ChronTree $T$ is the proof of presence in $T$ of the last
inserted data of $T'$, \emph{i.e.} $d_{N'}$ when $\Seq{T'} =
[d_1,\ldots, d_{N'}]$. With this proof and the sizes of both trees, we
can reconstruct the label of the root $T$ but also the label of the
root of $T'$ as means to verify the proof of extension. Note that when
$N' = 2^{k}$ for some $k$, it implies that the tree $T'$ is perfect
and so the label of the root of $T'$ is also a label of a node in $T$.
Therefore, to reconstruct the label of the root of $T$, we only need a
fragment of the proof of presence of $d_{N'}$ in $T$.

\begin{example}
\label{ex:verif extension}
Coming back to Example~\ref{ex: proof of extension CT}, consider the
bitstrings $h_b =
\hfun(\hfun(\hfun(d_1,d_2),\hfun(d_3,d_4)),\hfun(d_5,d_6))$ and $h_a =
\hfun(\hfun(d_1,d_2),d_3)$. $seq$ proves the extension of $h_a$ of
size $3$ into $h_b$ of size $6$. Figure~\ref{fig:example verification
  proof of extension CT} is the graphical representation of the
verification of $seq$ given $h_a$ and $h_b$. In particular,
$(\ell\cdot r \cdot \ell, [d_4,\hfun(d_1,d_2),\hfun(d_5,d_6)])$ proves
the presence of $d_3$ in $h_b$ and $(r, [\hfun(d_1,d_2)])$ proves the
presence of $d_3$ in $h_a$.
\end{example}

\subsection*{Ordered data structure}
\label{sec:LexTree}

The ordered data structure is implemented as the combination of a
binary search tree and a Merkle tree. The idea is that we can regroup
all the information about a subject into a single node of the binary
search tree, and while being able to efficiently generate and verify
the proof of presence. We consider a total order on bitstrings denoted
$\leq$. This order could be the lexicographic order in the ASCII
representations but it could be any other total order on
bitstrings. The implementation is called $LexTree$.

A \emph{LexTree} $T$ is a binary search tree over pairs of bitstrings
\begin{itemize}
\item for all two pairs $(d,h)$ and $(d',h')$ of bitstrings in $T$,
  $(d, h)$ occurs in a node left of the occurrence of $(d',h')$ if and
  only if $d \leq d'$ lexicographically;

\item for all nodes $n \in T$, $n$ is labelled with the pair
  $(d,\hfun(d, h_\ell,h_r))$ where $d$ is some bistring and $(d_\ell,
  h_\ell)$ (resp. $(d_r, h_r)$) is the label of its left child
  (resp. right child) if it exists; or the constant $\Null$ otherwise.
\end{itemize}

Note that contrary to a ChronTree, the same set of data can be
represented by different LexTrees depending on how the tree is
balanced. To avoid this situation, we assume that there is a
pre-agreed way for balancing trees.





\begin{example}
\label{ex:LexTrees}
The tree in Figure~\ref{fig:example LexTree} is an example of LexTree
where $d_1 \leq d_2 \leq \ldots \leq d_{12}$.
\end{example}

\begin{example}
\label{ex: proof of presence LT}
Consider the LexTree $T$ of Figure~\ref{fig:example LexTree}. The
proof of presence of $d_8$ in $T$ is the tuple
$\LTPoP{h_\ell}{h_r}{seq_d}{seq_h}$ where:
\begin{itemize}
\item $h_\ell = \hfun{_7}$ and $h_r = \hfun{_9}$; and
\item $seq_d = [d_{10}, d_6]$
\item $seq_h = [ \hfun(d_{12},\hfun{_{11}},\Null), \
  \hfun(d_4,\hfun(d_2,\hfun{_1},\hfun{_3}), \hfun{_5})]$
\end{itemize}
\end{example}

Like in ChronTrees, verifying the proof of presence of some data $d$
in a LexTree $T$ consists of reconstructing the hash value of the root
of $T$.

\begin{example}
\label{ex: proof of absence LT}
Consider the $T_c$ of Figure~\ref{fig:example LexTree}. Consider some
data $d$ such that $d_7 \leq d \leq d_8$. The proof of absence of $d$
in $T_c$ is the tuple $\LTPoAbs{\Null}{\Null}{seq_d}{seq_h}$ where:
\begin{itemize}
\item $seq_d = [d_7,d_8,d_{10}, d_6]$
\item $seq_h = [\hfun{_9},\ \hfun(d_{12},\hfun{_{11}},\Null), \
  \hfun(d_4,\hfun(d_2,\hfun{_1},\hfun{_3}), \hfun{_5}) ]$
\end{itemize}
\end{example}

\end{document}